\documentclass[usenatbib]{mn2e}

\usepackage{graphicx}
\usepackage{graphics}
\usepackage{amsmath}
\usepackage{color}
\usepackage{url}
\usepackage[caption=false]{subfig}
\usepackage{hyperref}

\newcommand{\hmpc}{$h^{-1}\ \rm{Mpc}$}
\newcommand{\hgpc}{$h^{-1}\ \rm{Gpc}$}
\newcommand{\mhmpc}{h^{-1}\ \rm{Mpc}}
\newcommand{\mhgpc}{h^{-1}\ \rm{Gpc}}
\newcommand{\xips}{$\xi(\pi,\sigma)$}
\newcommand{\xipzero}{$\xi(\pi=0,\sigma)$}
\newcommand{\Mtwel}{$M_{12}$}
\newcommand{\Msun}{$h^{-1}\ M_\odot$}
\newcommand{\lya}{Ly-$\alpha$}
\newcommand{\bdla}{b_{\rm{DLA}}}
\newcommand{\kms}{\rm{km}\ \rm{s}^{-1}}

\def\aap{A\&A}
\def\aj{AJ}
\def\apj{ApJ}
\def\apjl{ApJL}
\def\apjs{ApJS}
\def\jcap{JCAP}
\def\mnras{MNRAS}
\def\prd{PhRvD}

\begin{document}

\title[Cross-Correlations with LyMAS]{Modeling Lyman-$\alpha$ Forest
Cross-Correlations with LyMAS}

\author[Lochhaas et al.]{Cassandra Lochhaas$^{1}$, David H.~Weinberg$^{1}$,
S\'ebastien Peirani$^{2}$, Yohan Dubois$^{2}$,
\newauthor St\'ephane Colombi$^{2}$, J\'er\'emy Blaizot$^{3}$, Andreu Font-Ribera$^{4,5}$, Christophe Pichon$^{2,6}$,
\newauthor Julien Devriendt$^{2}$ \\
 $^{1}$ Department of Astronomy and CCAPP, The Ohio State University, 
 140 West 18th
 Avenue, Columbus, OH 43210, USA \\
$^{2}$ Institut d'Astrophysique de
 Paris, Sorbonne Universiti\`{e}s, UPMC Univ Paris 6 et CNRS, UMR 7095, 98
 bis Bd Arago, 75014 Paris, France \\
$^{3}$ Universit\'{e} de Lyon, Lyon,
 F-69003, France \\
$^{4}$ Lawrence Berkeley National Laboratory, 1
 Cyclotron Road, Berkeley, CA, USA \\
$^{5}$ Kavli IPMU (WPI), UTIAS, The University of Tokyo, Kashiwa, Chiba 277-8583, Japan \\
$^{6}$ Institute of Astronomy, Cambridge University, Madingley Road, Cambridge, UK}

\maketitle

\begin{abstract}
We use the \lya\ Mass Association Scheme (LyMAS; Peirani
et al.\ 2014) to predict cross-correlations at $z=2.5$ between dark matter
halos and transmitted flux in the \lya\ forest, and compare to 
cross-correlations measured for quasars and damped \lya\
systems (DLAs) from the Baryon Oscillation Spectroscopic Survey (BOSS) by
Font-Ribera et al.\ (2012, 2013). We calibrate LyMAS using
Horizon-AGN hydrodynamical cosmological simulations of a $(100\ \mhmpc)^3$
comoving volume. We apply this calibration
to a $(1\ \mhgpc)^3$ simulation realized with $2048^3$ dark matter
particles. In the 100 \hmpc\ box, LyMAS
reproduces the halo-flux correlations computed from the full hydrodynamic
gas distribution very well. In the 1
\hgpc\ box, the amplitude of the large scale cross-correlation tracks the halo
bias $b_h$ as expected.
We provide empirical fitting functions that describe our numerical results.
In the transverse separation bins used for the BOSS analyses, LyMAS
cross-correlation predictions follow linear theory accurately down to small
scales. Fitting the BOSS measurements
requires inclusion of random velocity errors; we find best-fit RMS velocity 
errors of $399\ \kms$ and $252\ \kms$
for quasars and DLAs, respectively. We infer bias-weighted mean halo masses of
$M_h/10^{12}\ h^{-1}M_\odot=2.19^{+0.16}_{-0.15}$ and $0.69^{+0.16}_{-0.14}$
for the host halos of quasars and DLAs, with $\sim 0.2$ dex systematic uncertainty 
associated with redshift evolution, IGM parameters, and selection of data fitting range.
\end{abstract}

\begin{keywords}
large-scale structure of Universe
\end{keywords}

\section{Introduction}
\label{sec:intro}

The enormous high-redshift quasar sample observed by the Baryon Oscillation
Spectroscopic Survey \citep[BOSS;][]{Dawson2013} of SDSS-III \citep[SDSS =
Sloan Digital Sky Survey]{Eisenstein2011} has enabled the first empirical
studies of the large scale 3-dimensional auto-correlations of transmitted
flux in the \lya\ forest \citep[e.g.,][]{Slosar2011} and of the
cross-correlations between the \lya\ forest and other tracers of structure
such as damped \lya\ systems \citep[DLAs; see][hereafter FR12]{FR2012} and
quasars \citep[hereafter FR13]{FR2013}. Auto-correlation analyses allow
high-precision measurements of the angular diameter distance and Hubble
parameter via baryon acoustic oscillations \citep[BAO;
see][]{Busca2013,Slosar2013,Delubac2015}. Cross-correlation analyses
provide a complementary route to BAO measurements \citep{FR2014}, and they
provide novel constraints on the properties of dark matter halos that host
DLAs and quasars. On large scales, these correlations can be described by
linear theory with an effective bias factor $b_F$ and redshift-space
distortion parameter $\beta_F$ for the forest that depend on the underlying
cosmology and the physical parameters of the intergalactic medium
\citep[IGM; see, e.g.,\\][]{McDonald2003,Seljak2012,Arinyo2015}. 
Fully exploiting these
measurements, however, requires a theoretical description of \lya\ forest
clustering that extends to non-linear scales. In this paper we use the
\lya\ Mass Association Scheme \citep[LyMAS;][hereafter P14]{Peirani2014} to
predict the cross-correlation between dark matter halos and the \lya\
forest, and we use these predictions to model the measurements of DLA- and
quasar-forest cross-correlations by FR12 and FR13.

Ideally one would model \lya\ forest correlations using hydrodynamic
simulations that resolve the $\sim100$ kpc Jeans scale of the low density
IGM \citetext{see \citealt{Cen1994,Zhang1995,Hernquist1996,Miralda1996} for
pioneering studies and \citealt{Lidz2010,Peeples2010,Borde2014,Arinyo2015}
for some recent examples}. However,
$\sim$Gpc$^3$ simulation volumes are needed to avoid artificial box size
effects and exploit the statistical precision achievable with BOSS, and
calculations with the required combination of volume and resolution are not
currently feasible. One approach to this challenge
\citep[e.g.,][]{Slosar2009} uses variants of the Fluctuating Gunn-Peterson
Approximation \citep[FGPA;][]{Gunn1965,Croft1998,Weinberg1998}, a
deterministic mapping between dark matter density and \lya\ flux. LyMAS
instead uses high-resolution hydrodynamic simulations to calibrate the
conditional probability distribution $P(F_s|\delta_s)$ for the transmitted
flux smoothed (in one dimension) on the scale of BOSS spectral resolution
given the redshift-space dark matter density contrast $\delta_s$ smoothed
(in three dimensions) on a similar scale. This conditional distribution is
applied to the smoothed dark matter density field of a larger volume, lower
resolution $N$-body simulation to determine the \lya\ forest flux along
skewers through the matter distribution. Additional steps are used to
impose pixel-to-pixel coherence of spectra along a given line of sight and
to ensure that the 1-d power spectrum and unconditional probability
distribution function (PDF) of flux matches that of the calibrating
hydrodynamic simulations (see P14).

P14 calibrated $P(F_s|\delta_s)$ using the Horizon-MareNostrum simulation,
a hydrodynamic simulation of a
(50 \hmpc)$^3$ comoving volume with WMAP1 cosmological parameters
\citep[$h\equiv H_0/100$ km s$^{-1}$ Mpc$^{-1}$]{Spergel2003}. They
applied this calibration to a $1024^3$ simulation of a (300 \hmpc)$^3$ cube with
a Gaussian dark matter (DM) smoothing scale $r_s=0.3$ \hmpc\ and to a
$1024^3$ simulation of a (1 \hgpc)$^3$ cube with $r_s=1.0$ \hmpc. In this
work we improve the calibration using Horizon-AGN and Horizon-noAGN, two
simulations of a (100 \hmpc)$^3$ volume adopting WMAP7 parameters
\citep{Komatsu2011}, one with and one without feedback from active galactic
nuclei (AGN), as described by \citet{Dubois2014}. For large volumes we
use one $2048^3$ simulation of a (1 \hgpc)$^3$ cube, with dark matter smoothing
$r_s=0.5$ \hmpc, and several $1024^3$ simulations of the same volume with
$r_s=1.0$ \hmpc. These ``upgrades" improve the accuracy and precision of
our predictions and give us a better handle on statistical uncertainties
and the potential impact of AGN feedback.

FR12 measured the cross-correlation \xips\ between DLAs and the \lya\
forest, as a function of line-of-sight 
separation $\pi$ in bins of transverse
separation $\sigma$, using $\sim60,000$ quasar spectra from the Data
Release 9 (DR9; \citealt{Ahn2012}) BOSS quasar sample
\citep{Ross2012,Paris2012}. They modeled these measurements using linear
theory over transverse scales $\sigma=1-60$ \hmpc\ (comoving), excluding
scales $r=(\sigma^2+\pi^2)^{1/2}<5$ \hmpc\ in all bins. Adopting \lya\
forest linear bias parameters $b_F=-0.168$, $\beta_F=1.0$ at effective
redshift $z=2.3$, they inferred a linear bias for DLAs of
$b_{\mathrm{DLA}}=2.17\pm0.2$ at $z=2.3$, corresponding to a characteristic
halo mass $M_h=4.5\times10^{11}$ \Msun\ (which yields the same bias factor
at this redshift). FR13 applied similar techniques to the
cross-correlation of DR9 quasars with the \lya\ forest, extending
measurements to $\sigma=80$ \hmpc\ and excluding scales $r<15$ \hmpc\ from
their linear fits in all bins. They inferred $b_q=3.64^{+0.13}_{-0.15}$
for the average bias factor of DR9 quasars, with a corresponding halo mass
$M_h=3.26\times10^{12}$ \Msun. Similar analyses using the BOSS DR12
\citep{Alam2015} \lya\ forest, DLA, and
quasar samples are underway.

In \S\ref{sec:sim}, we give an explanation of how LyMAS is calibrated on a
small-volume hydrodynamic simulation and then applied to large-scale dark
matter simulations, and we briefly examine the influence of AGN feedback on
the \lya\ forest. In \S\ref{sec:xcorr}, we present our results for the
dependence of the cross-correlation on various properties and provide
comparisons of our calculations to linear theory predictions for the
correlation. In \S\ref{sec:LvsH}, we analyze the accuracy of LyMAS for
predicting cross-correlations by examining how different smoothing lengths,
box sizes, and particle numbers affect the correlation. We fit a simple,
three-parameter model to the cross-correlation in \S\ref{sec:Mh} to
determine the dependence of the strength, width, and offset of the
correlation on halo mass. We compare our results to linear theory
predictions for the \lya\ forest auto-correlation and halo-\lya\
cross-correlation in \S\ref{sec:linth}. We compare our calculations for
halo-forest cross-correlations to measured cross-correlations of DLAs and
quasars with the \lya\ forest in \S\ref{sec:obs}. We present our
conclusions in \S\ref{sec:con}.

\section{Simulations}
\label{sec:sim}

A detailed account of how LyMAS operates is given by P14, but we summarize
the main points and present some changes since the original paper here.
LyMAS uses PDFs for the \lya\ forest flux that are conditioned on the
underlying dark matter density to predict the forest flux in a large-volume
DM simulation. These conditional PDFs are determined from a full
hydrodynamics and $N$-body simulation using the RAMSES code
\citep{Teyssier2002} in a periodic box with side length of 100 \hmpc\ (50
\hmpc\ in P14), adopting WMAP7 cosmological parameters (WMAP1 in P14):
$\Omega_m=0.272$, $\Omega_\Lambda=0.7284$, $\Omega_b=0.045$, $h=0.704$,
$\sigma_8=0.81$, $n_s=0.967$. The conditional PDFs are then
applied to large-scale $N$-body simulations created using Gadget2
\citep{Springel2005}, from which pseudo-spectra are extracted.

The smaller hydrodynamics simulation includes physical processes such as
metal-dependent cooling, photoionization and 
heating from a UV background, supernova feedback,
and metal enrichment. To calibrate the conditional PDFs, about one million
lines of sight are extracted from this simulation, and the optical depth of
\lya\ absorption is calculated based on the neutral hydrogen density along
the line of sight. The spectra are smoothed with a 1-d Gaussian of
dispersion $0.696$ \hmpc, 
equivalent to BOSS spectral resolution at
$z\approx2.5$. DM skewers that correspond to the spectra are also
extracted from this simulation, and the DM is smoothed three-dimensionally
with several smoothing scales $r_s$. The optical depth along each spectrum
is converted to \lya\ forest flux by $F=e^{-\tau}$, and PDFs for the values
of the flux conditional on DM overdensity can thus be determined. 

LyMAS operates on the assumption that the flux in one pixel is not correlated
with the flux in a separate pixel \emph{except} through correlations in the
underlying dark matter. The flux in pixel 1 is drawn from the conditional
distribution $P(F_1|\delta_1)$ without reference to the value $F_2$ drawn
from $P(F_2|\delta_2)$ in pixel 2, but $F_1$ and $F_2$ are correlated
because of the correlation between $\delta_1$ and $\delta_2$. This {\it
ansatz}, that the relation between flux and matter density is ``locally
stochastic,'' can be expressed by the equation 
\begin{equation}
P(F_1,F_2|\delta_1,\delta_2) = P(F_1|\delta_1)P(F_2|\delta_2)~.
\end{equation}
Along each individual
line of sight, coherence among neighboring pixels is imposed by
generating a ``percentile field" that governs draws from $P(F_s|\delta_s)$,
rather than drawing each pixel value independently. The correct 1-d flux
power spectrum is imposed, on average, by multiplying line-of-sight Fourier
modes by numerically calibrated correction factors. Except for the change
in the calibrating simulations themselves, our procedures here are the same
as those presented by P14.

\begin{figure*}
\begin{minipage}{175mm}
\centering
\includegraphics[width=.48\linewidth]{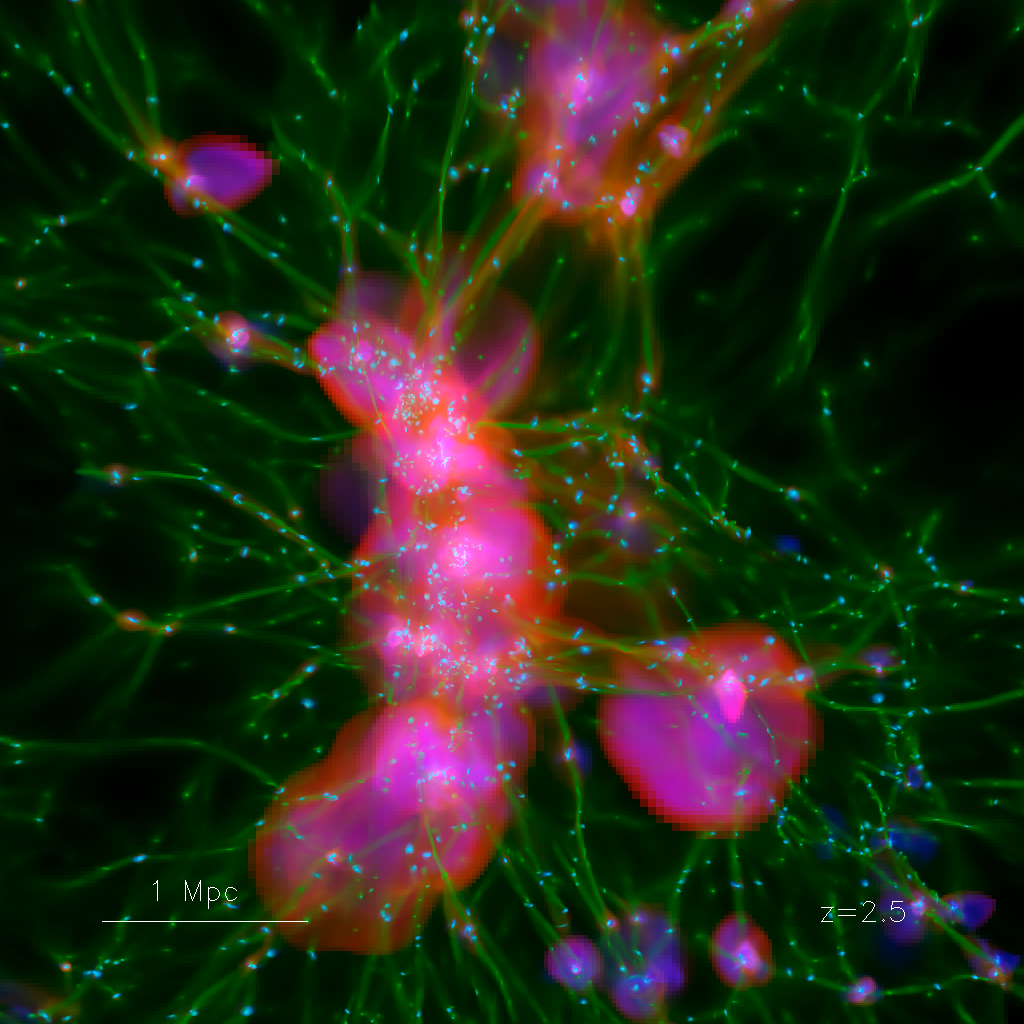}
\includegraphics[width=.48\linewidth]{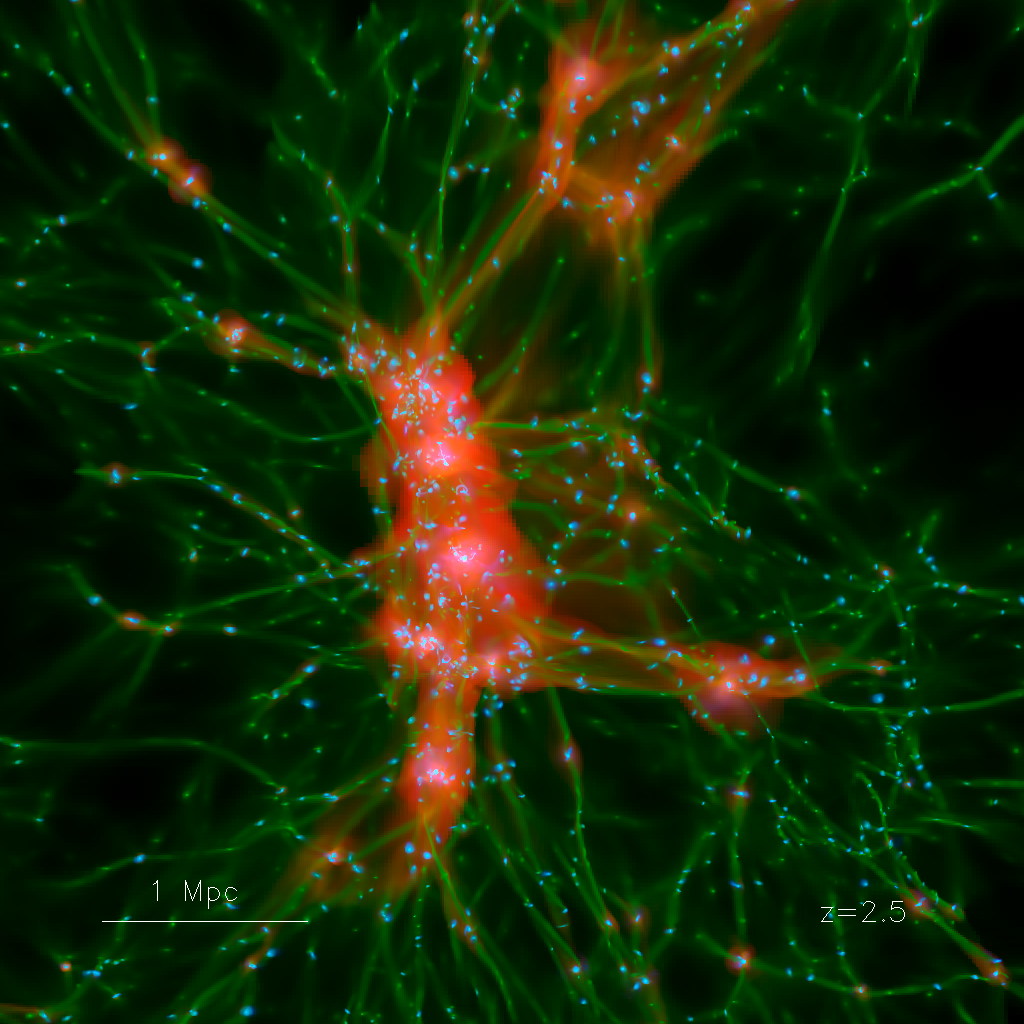}
\caption{The same slice
through a hydrodynamics simulation that includes AGN feedback (left) and
one that does not (right). Blue colours trace regions of high gas
metallicity, red colours trace regions of high gas temperature, and green
colours trace regions of high gas density.}
\label{fig:AGN_slices}
\end{minipage}
\end{figure*}

We apply LyMAS to DM simulations with side
lengths 300 \hmpc\ and 1 \hgpc, the same sizes studied in P14, but with
cosmological parameters matched to those of the hydro simulations. Our
primary simulations have $1024^3$ particles for 300 \hmpc\ and $2048^3$
particles for 1 \hgpc, and we analyze them with DM smoothing lengths of 0.3
\hmpc\ and 0.5 \hmpc, respectively, equal to the initial mean interparticle
separation. We also carried out a $1024^3$ simulation of the 1 \hgpc\ box
with the same initial conditions and two $1024^3$ simulations of 1 \hgpc\
boxes with independent initial fluctuations, all analyzed with a 1.0 \hmpc\
DM smoothing length. We apply all three smoothing lengths, $r_s=0.3,\
0.5,\ 1.0$ \hmpc, to the 100 \hmpc\ hydro simulation to calibrate
conditional PDFs and to test the accuracy of LyMAS. 

The hydrodynamical cosmological simulation Horizon-AGN is described in
detail by \citet{Dubois2014}. The initial grid for the gas distribution is
$1024^3$, with a comoving grid cell of 0.25 \hmpc, and adaptive mesh
refinement (AMR) is used up to a maximum resolution (for hydrodynamics and
gravity) of 1 proper kpc. The dark matter distribution is represented by
$1024^3$ particles, with effective 
gravitational softening set by the cloud-in-cell
interpolation of the particle mass distribution
onto the adaptive grid. Fig.~\ref{fig:AGN_slices} shows a
slice through a simulation that includes AGN feedback (Horizon-AGN, left)
and the corresponding slice through a simulation without AGN feedback
(Horizon-noAGN, right). While the gas density field (green) is similar in
both simulations, regions of high temperature (red) and high metallicity
(blue) are more extended in the AGN feedback run. AGN feedback has a
considerable impact on the mass function and morphologies of the simulated
galaxy population \citep{Dubois2013}, but the impact on conditional flux
PDFs is small, and (as we show below) the impact on halo-flux correlations
is negligible. When extracting \lya\ forest spectra from the simulated gas
distribution, the UV background intensity is chosen to give a mean
transmitted \lya\ forest flux $\bar F = \left<e^{-\tau}\right>=0.795$,
matching the metal-corrected $z=2.5$ value measured from high-resolution
spectra by \cite{Faucher-Giguere2008}.

We identify DM halos using a friends-of-friends algorithm \citep{Davis1985}
with a linking length of $0.15$. Then we compute the center of mass of
each object. Once we have pseudo-spectra determined by LyMAS and the
positions and masses of DM halos in the $N$-body simulations, we calculate
the cross-correlation by randomly selecting a large number of
pseudo-spectra and DM halos from each volume. We bin the DM halos by mass,
and for each halo we bin the pseudo-spectra around it in transverse and
line-of-sight separations $\sigma$ and $\pi$. We calculate the
cross-correlation $\xi$ as 
\begin{equation}
\xi(\pi,\sigma)=\frac{\langle F(\pi, \sigma) \rangle}{\bar F} - 1 ~,
\label{eq:xcorr}
\end{equation} 
where $F(\pi, \sigma)$
is the transmitted flux fraction ($F=1$ for no absorption by the \lya\
forest, $F=0$ for complete \lya\ absorption) at line-of-sight separation
$\pi$ and transverse separation $\sigma$ from the center of mass of the
halo, where $\bar F$ is the mean \lya\ forest transmitted flux in the
entire box. The average of the flux (indicated by $\langle\rangle$) is performed 
over all pixels that fall into the halo mass, $\pi$, and $\sigma$ bin.
Equation (\ref{eq:xcorr}) is equivalent to the cross-correlation between the halo 
number density field and the flux fluctuation field
\begin{equation}
\delta_F=\frac{F-\bar F}{\bar F}~.
\end{equation}
We also calculate the auto-correlation as 
\begin{equation}
\xi(r,\mu)=\langle\delta_{F1}\delta_{F2}\rangle ~,
\end{equation} 
where
$\delta_F=(F-\bar F)/\bar F$, $r$ denotes the comoving separation between
the pixels where $F_1$ and $F_2$ are measured, and $\mu=\pi/r$ denotes
the cosine of the angle from the line of sight between the two pixels. All calculations
are done in redshift space, incorporating the simulated peculiar velocities
of the gas, DM particles, and halos.

\section{LyMAS Predictions for Halo-Forest Cross-Correlations}
\label{sec:xcorr}

In this section, we assess the accuracy of LyMAS as a tool for predicting
halo-forest cross-correlations by comparing results for a variety of
simulations. After demonstrating the expected accuracy of our largest
volume simulation, we fit a model to the numerical cross-correlation
results and analyze its dependence on halo mass. We also compare the LyMAS
predictions to linear theory to infer where nonlinear effects have a
significant impact on auto- and cross-correlations.

\subsection{Accuracy of LyMAS}
\label{sec:LvsH}

\begin{figure*}
\begin{minipage}{175mm}
\centering
\includegraphics[width=\linewidth]{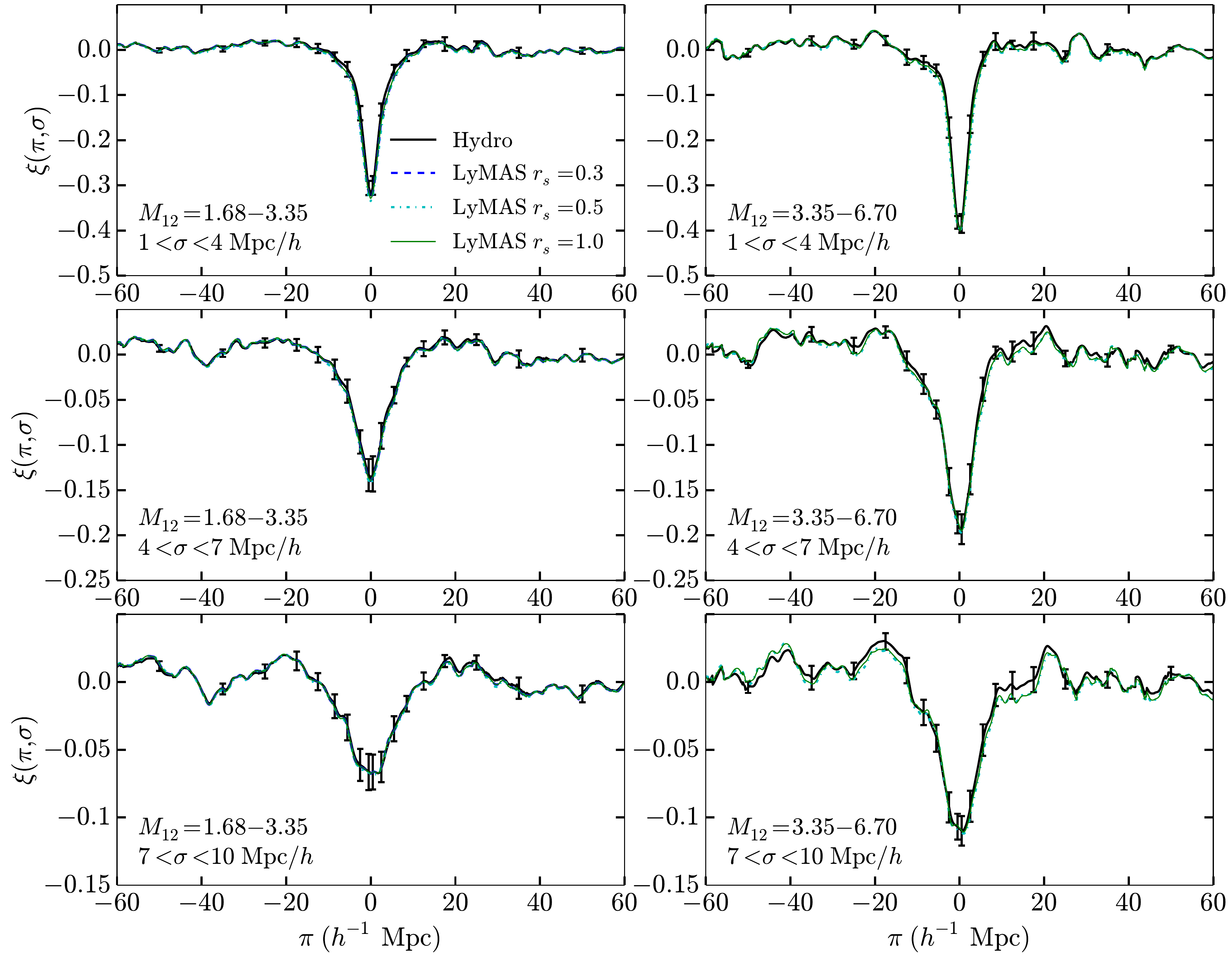}
\caption{The cross-correlation between dark matter halos and \lya\ forest
flux calculated from true gas spectra (black solid) and from LyMAS applied
to the matter distribution with 0.3 \hmpc\ 3-d dark matter smoothing (blue
dashed), 0.5 \hmpc\ smoothing (cyan dot-dashed), or 1.0 Mpc \hmpc\
smoothing (green solid) in the (100 \hmpc)$^3$ simulation. Rows show
transverse separation bins $\sigma=1-4,\ 4-7$, and $7-10$ \hmpc, and
columns show dark matter halo mass bins \Mtwel\ $=1.68-3.35$ and
$3.35-6.70$. Similar agreement holds in other mass and separation bins. 
Error bars are computed from the standard deviation 
of the mean among 16 subvolumes.}
\label{fig:LyMAS_vs_hydro}
\end{minipage}
\end{figure*}

Fig.~\ref{fig:LyMAS_vs_hydro} compares the halo-forest cross-correlation
\xips\ computed from the gas distribution of the noAGN hydrodynamic
simulation (see \S\ref{sec:sim}) to the cross-correlations predicted by
applying LyMAS to the smoothed DM density fields of the same simulation
with $r_s=0.3,\ 0.5,$ or $1.0$ \hmpc. Each panel shows the correlation as
a function of line-of-sight separation $\pi$ between the center of mass of
each halo and pixels in the spectra. The three rows show three different
halo-pixel transverse separation bins, $\sigma=1-4$ \hmpc, $4-7$ \hmpc,
and $7-10$ \hmpc, and the two columns show two different mass bins for the
halos, \Mtwel\ $\equiv M_h/10^{12}$ \Msun\ $=1.68-3.35$ and $3.35-6.70$.
As expected, the strength of the cross-correlation decreases with
increasing transverse or line-of-sight separation ($\sigma$ and $\pi$,
respectively) and increases with increasing halo mass \Mtwel. The four
curves are nearly indistinguishable in every panel, and we find similar
results for other transverse separation and halo mass bins. We conclude
that halo-forest cross-correlations predicted by LyMAS at BOSS spectral
resolution should be virtually identical to those that would be predicted
by a full high-resolution hydrodynamic simulation of the same volume, for
DM smoothing lengths up to $r_s=1.0$ \hmpc.

For representative statistical error bars, we divide the (100 \hmpc)$^3$
box into 16 subvolumes, with transverse size 25 \hmpc\ $\times$ 25 \hmpc\
and line-of-sight size 100 \hmpc. We calculate the cross-correlation,
\xips, in each of the 16 subvolumes and compute the error on the mean by
first finding the mean values in bins of $\pi$ (using the same bins as for
$\sigma$), then computing the error in each ($\pi,\sigma$) bin as the
standard deviation of the mean among the 16 subvolumes in the bin. We
include halo-pixel pairs that cross the subvolume divisions and assign
these pairs to the subvolume that contains the center of mass of the DM
halo. We always use the global mean flux $\bar{F}$ when calculating \xips,
instead of the mean flux within each subvolume. This procedure for
calculating errors may underestimate the statistical error we would find
from the dispersion among multiple independent simulations, or from
multiple (100 \hmpc)$^3$ volumes in a larger simulation, because it
underestimates the impact of coherent structure induced by large-scale
Fourier modes. The multiple curves in Fig.~\ref{fig:LyMAS_vs_hydro} agree
to much better than our estimated statistical errors in any case because
they are all derived from the same underlying full-cube matter
distribution.

\begin{figure*}
\begin{minipage}{175mm}
\centering
\includegraphics[width=\linewidth]{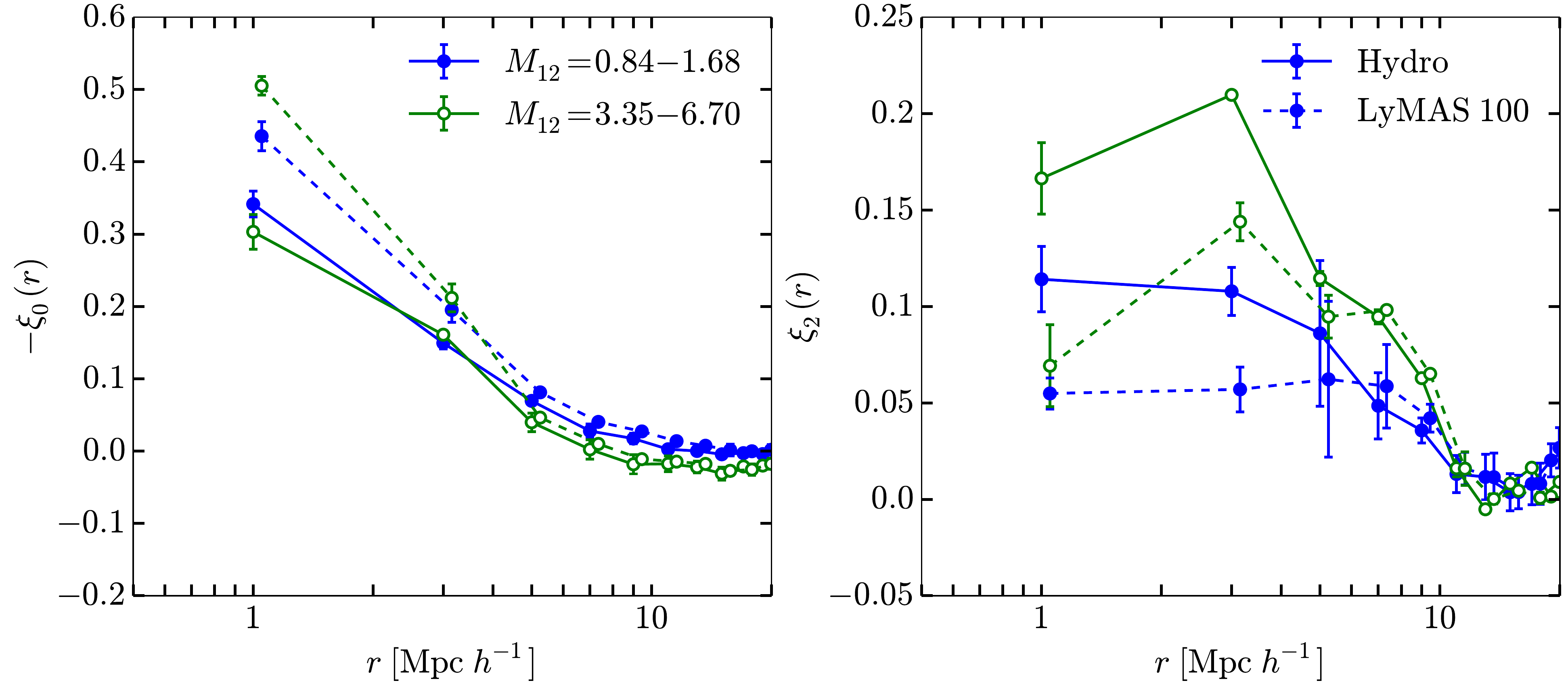}
\caption{The monopole (left) and quadrupole (right) of the halo-flux 
cross-correlation in the hydro box (solid) and for the LyMAS predictions 
applied to the (100 \hmpc)$^3$ dark matter simulation with 
$r_s=0.3$ \hmpc\ smoothing (dashed, horizontally offset for clarity). 
Halo mass bins \Mtwel\ $=0.84-1.68$ (filled circles) and $3.35-6.70$ 
(open circles) are shown. Error bars are computed from the standard deviation 
of the mean among 16 subvolumes.}
\label{fig:LyMAS_vs_hydro_mono_quad}
\end{minipage}
\end{figure*}

To further analyze how well the LyMAS predictions of the halo-flux 
cross-correlation agree with those from the hydrodynamic simulation, 
Fig.~\ref{fig:LyMAS_vs_hydro_mono_quad} shows the negative of the monopole 
(left panel) and the quadrupole (right panel) of the cross-correlation in
these simulations as functions of 3-d separation $r$ for two halo mass bins 
(see \S\ref{sec:linth} for a description of how the monopole and quadrupole 
are calculated). Small scale deviations between LyMAS and full hydro are more 
evident in this representation than in the \xips\ representation, which 
averages over a range of scales. LyMAS weakly over-predicts the strength of the 
monopole at $r\le 3$ \hmpc\ and under-predicts the strength of the quadrupole at 
$r\le6$ \hmpc, with a larger difference in the higher halo mass bin. 
On larger scales, the monopole and quadrupole of the LyMAS cross-correlation 
match the hydro correlation to well within the error bars.

\begin{figure*}
\begin{minipage}{175mm}
\centering
\includegraphics[width=\linewidth]{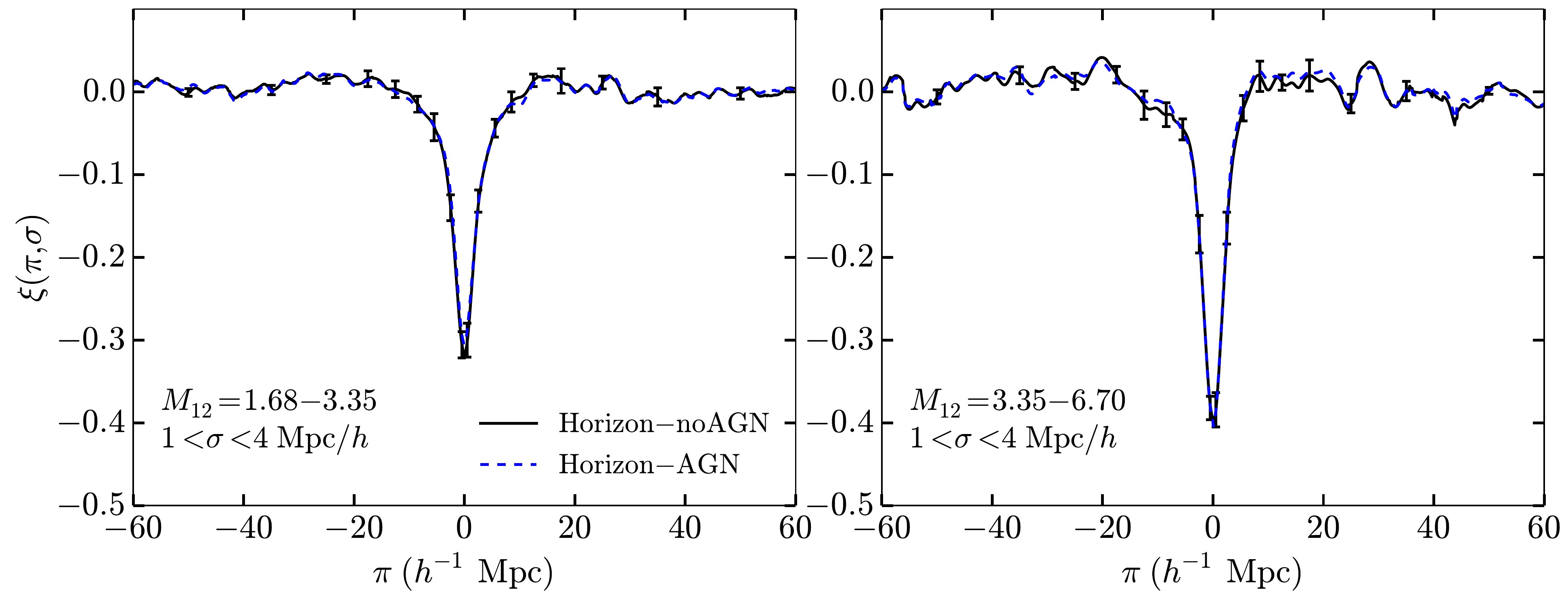}
\caption{The cross-correlation between dark matter halos and \lya\ forest
flux calculated in the (100 \hmpc)$^3$ hydrodynamic simulations using the
true gas spectra either including (blue dashed) or not including (black solid) AGN
feedback. Only the $\sigma=1-4$ \hmpc\ separation bin and \Mtwel\
$=1.68-3.35$ and $3.35-6.70$ mass bins are shown, but the near-perfect
agreement holds in other mass and separation bins. Error bars are computed 
from the standard deviation of the mean among 16 subvolumes.}
\label{fig:AGN}
\end{minipage}
\end{figure*}

As AGN feedback plays an important role in the temperature structure in
high density regions (see Fig.~\ref{fig:AGN_slices}), we examine its impact
on the halo-forest cross-correlation. Fig.~\ref{fig:AGN} compares the
correlation computed from the true gas spectra in the two hydro
simulations, with and without AGN feedback (as described in
\S\ref{sec:sim}), both in a (100 \hmpc)$^3$ volume. Only the $\sigma=1-4$
\hmpc\ transverse separation bin and the \Mtwel\ $=1.68-3.35$ and
$3.35-6.70$ mass bins are shown, but results are similar in other $\sigma$
and \Mtwel\ bins. There is virtually no difference in the
cross-correlation when AGN feedback is included as compared to when it is
not. Previous studies have shown that AGN feedback can have a small (few
percent) impact on the unconditional PDF and 1-d power spectrum of \lya\
transmitted flux \citep{Viel2013}, and we find similar effects in our
simulations. However, while AGN
feedback in these simulations has a substantial effect on the properties of
massive galaxies \citep{Dubois2013}, the impact on halo-forest
cross-correlations is negligible at the level of the statistical errors in
our simulation predictions and the observational data. We find a similarly
negligible effect if we use the conditional PDFs from either the AGN or
noAGN hydro simulation as our basis for applying LyMAS to larger volume
$N$-body simulations. In the remainder of this paper, we use the noAGN
simulation as the source of these conditional PDFs, but none of our
conclusions would change if we used the AGN simulation instead.

\begin{figure*}
\begin{minipage}{175mm}
\centering
\includegraphics[width=\linewidth]{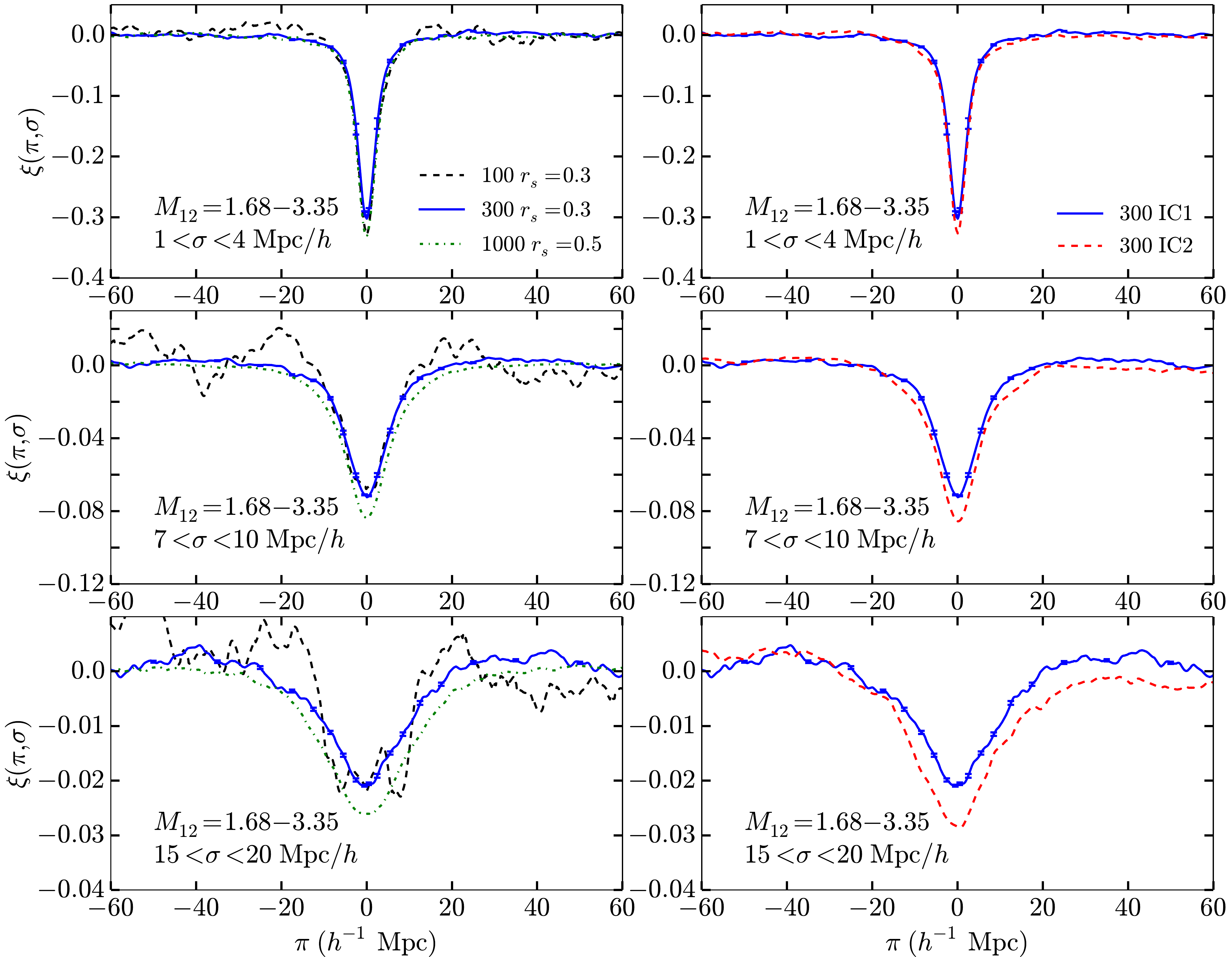}
\caption{Left column shows the halo-forest cross-correlation calculated by
applying LyMAS to (100 \hmpc)$^3$ (black dashed), (300 \hmpc)$^3$ (blue
solid), and (1 \hgpc)$^3$ (green dot-dashed) dark matter volumes, in
transverse separation bins $\sigma=1-4,\ 7-10,$ and $15-20$ \hmpc\ for
halos with \Mtwel\ $=1.68-3.35$. Right column shows the cross-correlation
from applying LyMAS to two different dark matter density distributions in
(300 \hmpc)$^3$ volumes, from simulations with the same cosmological parameters
but different initial conditions, in the same halo mass bin and transverse
separation bins as the left column.}
\label{fig:boxsize}
\end{minipage}
\end{figure*}

We next examine the effect on the cross-correlation of using larger volume
simulations. The left column of Fig.~\ref{fig:boxsize} shows the
correlation predicted by applying LyMAS to simulation volumes of (100
\hmpc)$^3$, (300 \hmpc)$^3$, and (1 \hgpc)$^3$ with 3-d dark matter
smoothing lengths of $r_s=0.3,\ 0.3$, and $0.5$ \hmpc, respectively. Only
the \Mtwel\ $=1.68-3.35$ halo mass bin is shown. Based on
Fig.~\ref{fig:LyMAS_vs_hydro}, we are confident that differences in the
calculated correlation are differences in the simulation volumes rather
than smoothing scale effects. In each of the three transverse separation
bins shown in the figure ($\sigma=1-4,\ 7-10$, and $15-20$ \hmpc), there
are significant differences in the strength of the correlation, most easily
seen at \xipzero. This difference is quite large in the larger transverse
separation bins, where the separation is a significant fraction of the
smaller simulation volumes. While these differences could be partly due to
random statistical fluctuations between the boxes, the weaker correlation
at large $\sigma$ in the smaller boxes is likely a systematic effect of the
absence of Fourier modes larger than the box size.

The right column of Fig.~\ref{fig:boxsize} compares the correlation
calculated by applying LyMAS to two different (300 \hmpc)$^3$ dark matter
distributions (solid and dashed) that differ only in the statistical
fluctuations of the initial conditions of the dark matter density. There
is a significant difference in the strength of the cross-correlation
between the two distributions, which means that (300 \hmpc)$^3$ is not a
large enough volume to overcome substantial statistical fluctuations in the
density distribution. For this reason, we use a (1 \hgpc)$^3$ box for all
following analysis of the cross-correlation. Fig.~\ref{fig:boxsize} also
makes the cautionary point that differences between simulation volumes can
sometimes be much larger than that estimated by our subvolume method,
though we show below that the subvolume error estimate appears reasonable
for the $(1\ \mhgpc)^3$ box at $\sigma > 7\ \mhmpc$.

\begin{figure*}
\begin{minipage}{175mm}
\centering
\includegraphics[width=\linewidth]{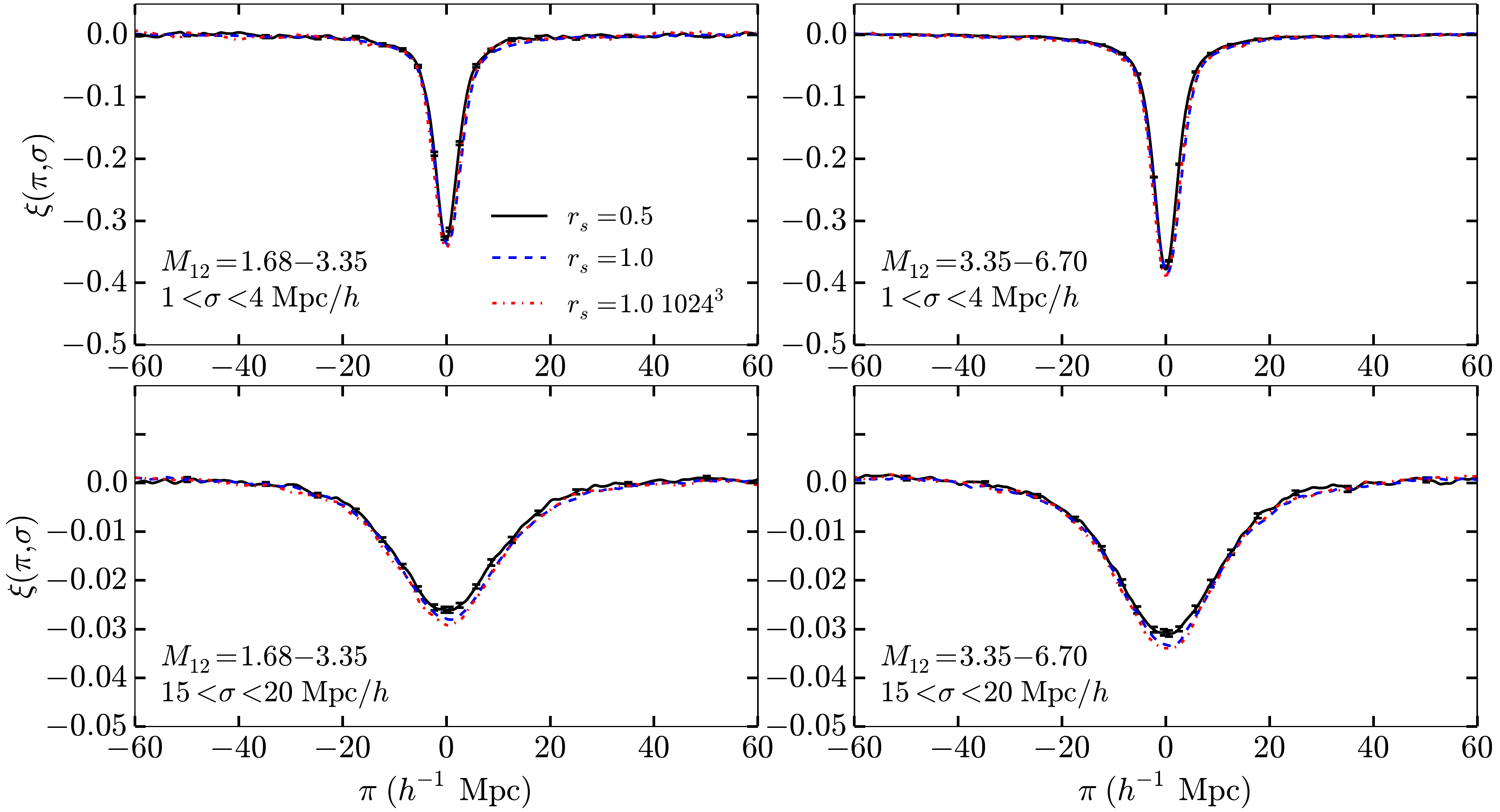}
\caption{The cross-correlation between dark matter halos and Ly-$\alpha$
forest flux from applying LyMAS to the (1 \hgpc)$^3$ dark matter simulation
with $2048^3$ particles for 0.5 (black solid) or 1.0 \hmpc\ (blue dashed) dark matter
smoothing scale, or for a (1 \hgpc)$^3$ DM simulation with $1024^3$
particles and 1.0 \hmpc\ smoothing scale (red dot-dashed). Left and right
columns show mass bins \Mtwel\ $=1.68-3.35$ and $3.35-6.70$; top and bottom
rows show transverse separation bins $\sigma=1-4$ \hmpc\ and $15-20$
\hmpc.}
\label{fig:resolution}
\end{minipage}
\end{figure*}

As a further examination of DM smoothing length and resolution effects,
Fig.~\ref{fig:resolution} shows the correlations in the (1 \hgpc)$^3$
volume using two different smoothing scales and two choices of $N$-body mass
resolution. Dashed and dot-dashed curves show LyMAS predictions for our
primary, $2048^3$-particle simulation and a simulation with the same
initial conditions but $1024^3$ particles, both for DM smoothing length
$r_s=1.0$ \hmpc. Results are indistinguishable, indicating that a
(Gaussian) DM smoothing length equal to the mean interparticle separation
yields converged results for the cross-correlation. Solid curves show
results for the $2048^3$ simulation with a smoothing length $r_s=0.5$
\hmpc, and in contrast to Fig.~\ref{fig:LyMAS_vs_hydro} we now see a slight
difference between $r_s=0.5$ \hmpc\ and $r_s=1.0$ \hmpc\ in the larger
($\sigma=15-20$ \hmpc) separation bin. P14 found that $r_s=0.3$ \hmpc\
yielded noticeably more accurate auto-correlation results than $r_s=1.0$
\hmpc, though Fig.~\ref{fig:LyMAS_vs_hydro} and similar tests we have
carried out for auto-correlations suggest that P14's results may have been
affected by the limited (50 \hmpc)$^3$ volume of their hydrodynamic
simulation. In what follows we use the $2048^3$ simulation with $r_s=0.5$
\hmpc\ for our main predictions but use $1024^3$ simulations with $r_s=1.0$
\hmpc\ to estimate statistical errors.

\subsection{LyMAS Predictions and Trends with Halo Mass}
\label{sec:Mh}

After experimenting with different functional forms, we have found that we
can achieve a visually good fit to the predicted cross-correlations in each
$\sigma$ bin for all of the mass ranges we have examined using a
3-parameter Lorentzian function 
\begin{equation}
\xi(\pi,\sigma)=-\alpha\frac{\gamma^2}{\pi^2+\gamma^2}+\Delta
\label{eq:Lorentz} 
\end{equation} 
where $\alpha$ is the correlation
strength at $\pi=0$ and $\gamma$ describes the width of the Lorentzian in
each $\sigma$ bin. The cross-correlation can be non-zero out to large
separations, and fits are significantly more accurate if we include a large
scale vertical offset $\Delta$ as an additional parameter. The purpose of these 
fits is partly to allow us to report our full set of numerical results in a 
compact form and partly to allow us to fit observational data with a continuous 
model rather than interpolate across the discrete mass bins used in our computation.

\begin{figure}
\centering
\includegraphics[width=\linewidth]{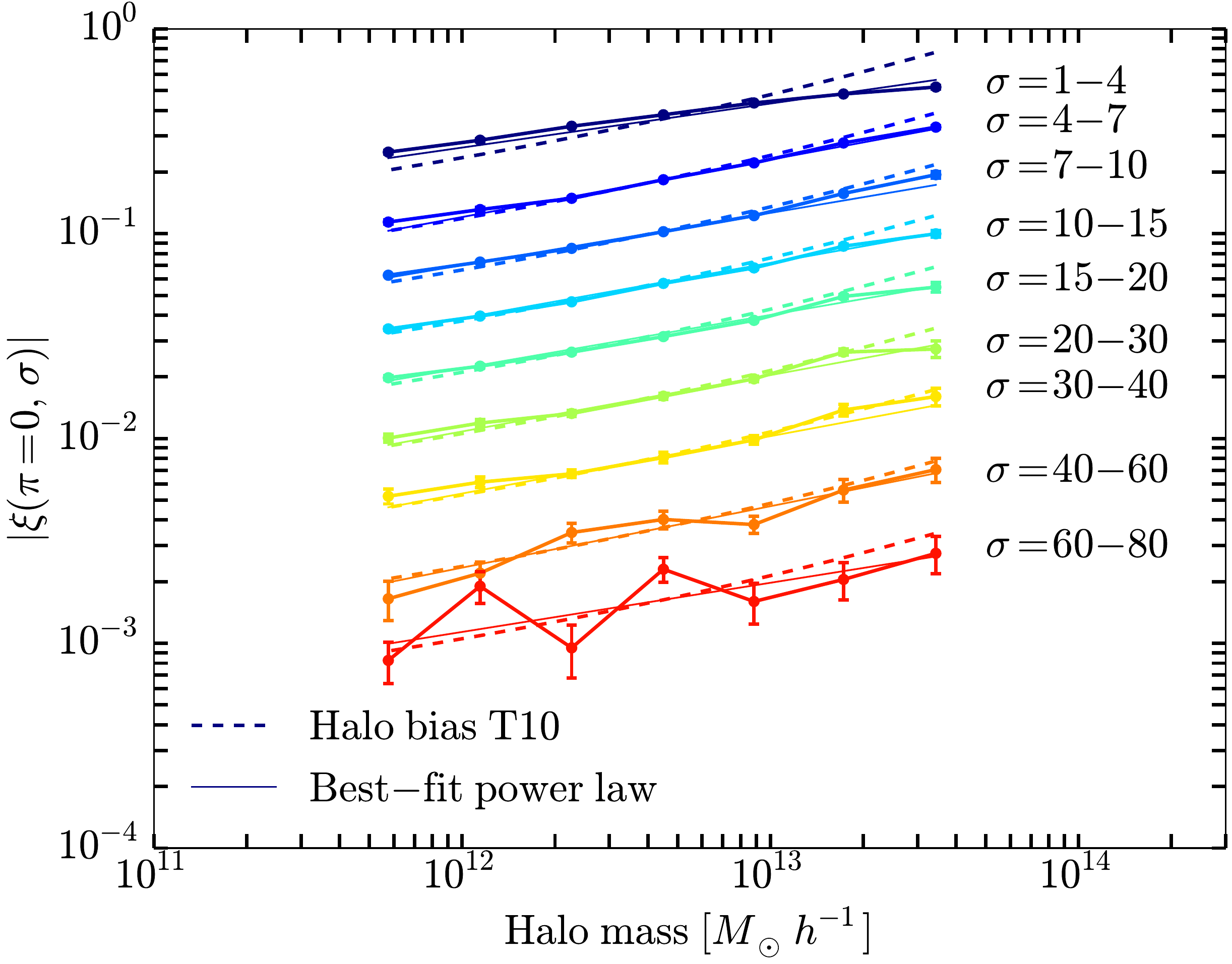}
\caption{The peak correlation $|$\xipzero$|$ of the best-fit Lorentzian
profiles as a function of average halo mass within each bin (thick solid
lines and points with errors) for each transverse separation bin (marked
next to each line on the right, increasing top to bottom). The dashed
lines show the dependence of the T10 halo bias on halo mass, scaled to a
height appropriate for each transverse separation bin. The thin solid
lines show a best-fit power law to each line.}
\label{fig:depvsmass}
\end{figure}

We fit the Lorentzian model of Eq.~\ref{eq:Lorentz} to the
cross-correlation predicted by LyMAS in each of the transverse separation
$\sigma$ and halo mass \Mtwel\ bins. Thick solid lines and points with
error bars in Fig.~\ref{fig:depvsmass} show the peak correlation strength,
$|$\xipzero$|$, as a function of the average halo mass within each bin, as
determined by the best-fit Lorentz profile. Errors on these points are the
dispersion in the best-fit Lorentz profile among the same 16 subvolumes
described in \S\ref{sec:LvsH}, found simply by fitting the same form to
\xips\ in each subvolume. (For each of these 16 fits, the errors on values
of \xips\ were the error on the mean of all pixels in the ($\pi$, $\sigma$)
bin.) As expected, the correlation strength increases with
decreasing $\sigma$ and with increasing halo mass. We fit a power-law of
the form 
\begin{equation}
\xi(\pi=0,\sigma) = A
\left(\frac{M_h}{4\times10^{12}\ h^{-1}M_\odot}\right)^m 
\label{eq:depth}
\end{equation} 
in each $\sigma$ bin; these fits are shown as the thin solid
lines, with best-fit parameters and errors shown in
Table~\ref{tab:mass_params}. The powers $m$ of halo mass are nearly
identical for all transverse separation bins except the smallest, meaning
that the mass dependence of the correlation strength is nearly the same at
all transverse separations. The best-fit scaling factor generally
decreases with increasing transverse separation. Errors on the parameters
of the fit are the dispersion of the best-fit parameter in the 16
subvolumes.

Dashed lines in Fig.~\ref{fig:depvsmass} show the \citet{Tinker2010}
(hereafter T10) halo bias formula as a function of halo mass at $z=2.5$,
scaled vertically to match the point at $\sim 4 \times 10^{12}$ \Msun\ for
each transverse separation bin. The predicted slope of the halo bias with
mass matches the measured slope of \xipzero\ with mass very well, except in
the smallest $\sigma$ bin, where our measured slope is lower than the halo
bias prediction. The T10 bias as a function of halo mass is not a perfect
power law over the mass range plotted here, and our highest mass points
tend to fall below this relation. This disagreement could be a consequence
of our finite box size, as even a (1 \hgpc)$^3$ box contains relatively few of
these extreme halos at this redshift (300 halos in the largest mass bin,
compared to $\sim1.58$ million in the smallest). As a representation of
our results, therefore, one can either take the power-law fits reported in
Table~\ref{tab:mass_params} or the T10 halo bias mass dependence normalized to the
reported fit amplitude at $M_h=4\times10^{12}$ \Msun\ (corresponding to
$\nu=2.75$ in the T10 formula for our adopted
cosmology and $z=2.5$).
Linear theory predicts that $\xi(\pi=0,\sigma)$ is nearly proportional
to $b_h$, but not exactly so because of the reduced redshift-space
distortion for higher halo bias (see equation~\ref{eq:linpk} below).

\begin{figure}
\centering
\includegraphics[width=\linewidth]{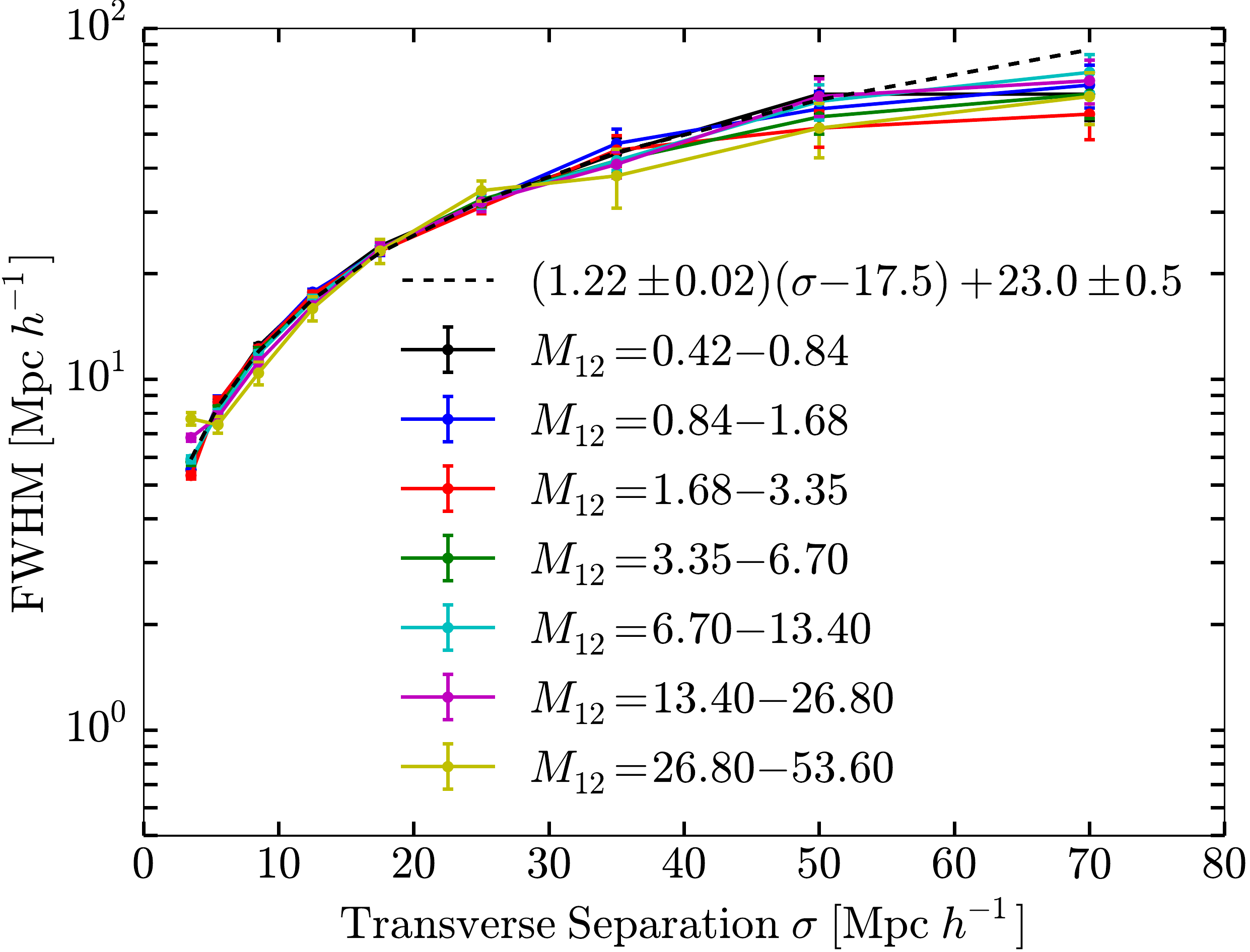}
\caption{The best-fit full width at half minimum $2\gamma$ of the
Lorentzian profiles as functions of the centers of each transverse
separation bin (solid lines). Different coloured lines show different halo
mass ranges, as listed in the plot legend. The black dashed line shows the
best-fit linear function to the width for all halo mass bins.}
\label{fig:FWHM}
\end{figure}

Fig.~\ref{fig:FWHM} plots the dependence of the best-fit width of the
Lorentz profiles on transverse separation and halo mass. The full width at
half-minimum, $2\gamma$, of the Lorentz profile has virtually no dependence
on halo mass, so we fit a single linear function to all bins simultaneously
around a pivot point at $\sigma=17.5$, obtaining 
\begin{equation}
\mathrm{FWHM}(\sigma) = 2\gamma = (1.22 \pm 0.02)(\sigma-17.5) + (23.0\pm0.5)~. 
\label{eq:FWHM} 
\end{equation} 
The relatively small errors on the
slope and intercept show this fitting function works quite well for the
relative width of the correlation in each transverse separation bin.
Again, errors are the error on the mean of the best parameters
fit to correlations in each of the 16 subvolumes.

\begin{figure*}
\begin{minipage}{175mm}
\centering
\includegraphics[width=\linewidth]{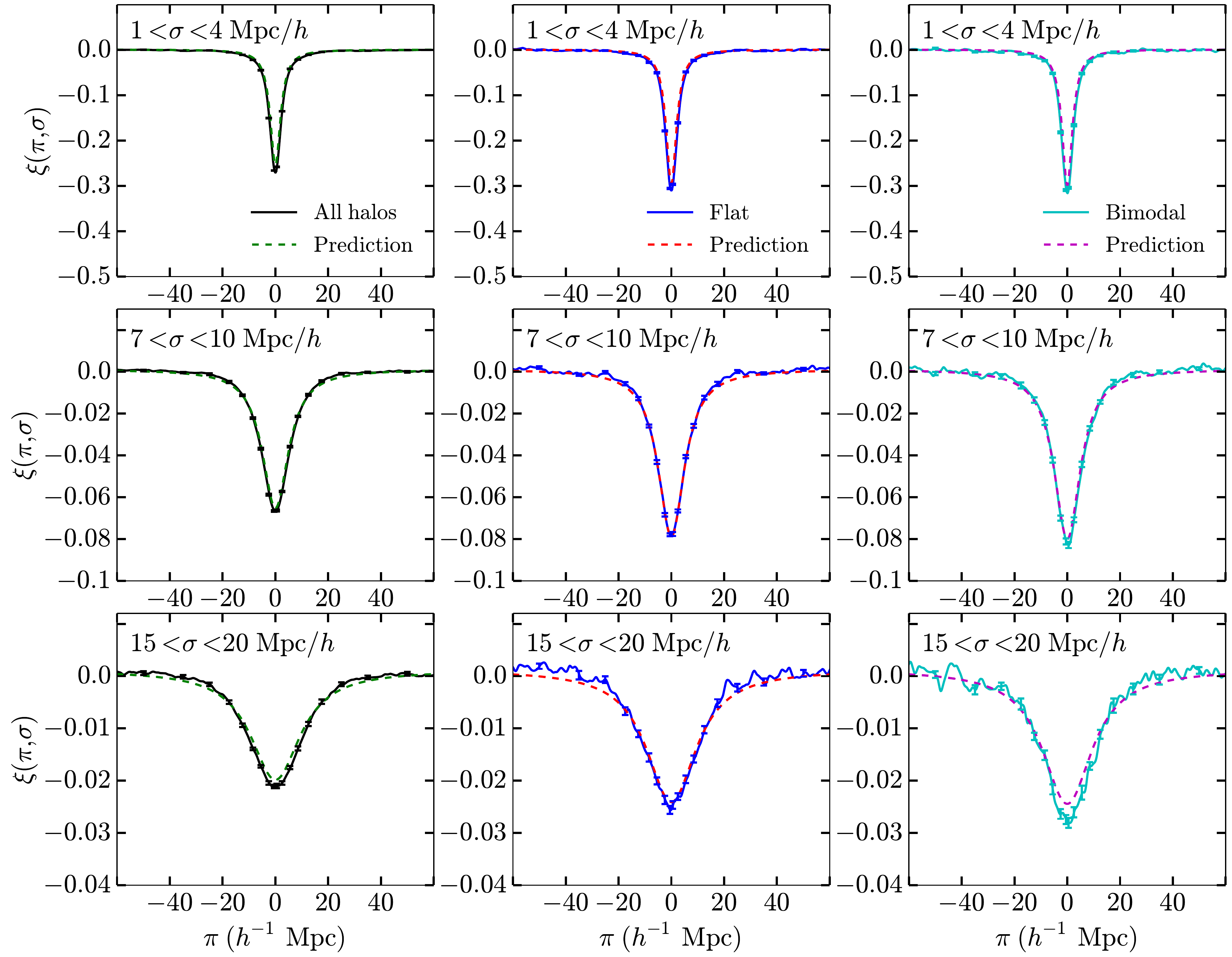}
\caption{The cross-correlation for three transverse separation bins (rows,
increasing top to bottom) and three halo mass distributions (columns)
plotted as solid lines, and a Lorentz profile with amplitude, width, and
offset based on the dependence on halo mass and transverse separation
determined by equations~(\ref{eq:depth}) and~(\ref{eq:FWHM}) and 
Tables~\ref{tab:mass_params} and~\ref{tab:delta} plotted as
dashed lines. Left column shows the cross correlation from a distribution
that reflects the mass distribution of all halos in the simulation, middle
column shows a flat distribution of halo masses, and right column shows a
bimodal distribution with equal numbers of high- and low-mass halos, and
none between.}
\label{fig:combmass}
\end{minipage}
\end{figure*}

We have so far considered correlations in narrow bins of halo mass, but
quasars and DLAs likely occupy a wide range of host halos. We now examine
the effect of a distribution of halo masses on the predicted correlation.
We choose halos within each of the lowest four halo mass bins, \Mtwel\
$=0.42-0.84,\ 0.84-1.68,\ 1.68-3.35,$ and $3.35-6.70$. We use three
different mass distributions: one in which the number of halos in each bin
reflects the mass distribution of all halos in the volume (``all halos",
the numbers are 158,068, 61,548, 21,150, and 6,363, respectively), one in
which the number of halos in each bin is the same (``flat", the number is
6,000), and one in which there are the same number of halos in the first
and last bins and no halos in the middle two bins (``bimodal", the numbers
are 6,000, 0, 0, and 6,000). Fig.~\ref{fig:combmass} shows the halo-flux
correlation for each of these distributions. Again, errors are from the 
subvolume method, as described in \S\ref{sec:LvsH}.

Also plotted in each panel of Fig.~\ref{fig:combmass} is a Lorentz profile
prediction for the correlation, with amplitude based on the best-fit power
laws in Fig.~\ref{fig:depvsmass} (dashed lines). Within any $\sigma$ bin,
the cross-correlation amplitude is approximately $\propto M_h^{0.272}$.
Therefore we expect, at least for linear bias, that the clustering of the
population will equal that of halos with the effective mass 
\begin{equation} 
M_{h,\mathrm{eff}} = \left[\frac{1}{N_{\mathrm{halos}}}
\sum_{i=1}^{N_{\mathrm{halos}}} M_{h,i}^{0.272}\right]^{1/0.272} ~, 
\label{eq:mass_eff}
\end{equation} 
where
$N_{\mathrm{halos}}$ is the total number of halos used in the correlation,
$M_{h,i}$ is the mass of each halo, and 0.272 is the weighted average of the
best-fit powers in the power-law dependence of the correlation on halo mass
for all but the smallest transverse separation bin, as given in
Table~\ref{tab:mass_params}. Equation (\ref{eq:mass_eff}) is equivalent to the 
condition
\begin{equation}
b(M_{h,\mathrm{eff}})=\frac{1}{N_{\mathrm{halos}}} \sum_{i=1}^{N_\mathrm{halos}} b(M_{h,i})
\label{eq:bias_eff}
\end{equation}
for a halo-bias relation $b\propto M_h^{0.272}$.
In Fig.~\ref{fig:combmass}, the amplitude of $|$\xipzero$|$ is computed from
this value of $M_{h,\mathrm{eff}}$ and the power-law fits in
Table~\ref{tab:mass_params},
and the width of the Lorentzian is determined
from the linear fit of equation~(\ref{eq:FWHM}). The prediction using our
best-fit parameters accurately matches the actual halo-flux correlation calculated
using LyMAS in all cases. In linear theory, the cross-correlation
strength should depend only on the number-weighted mean bias factor of
halos. This expectation holds for the LyMAS predictions even down to small
scales, indicating that bias-weighted cross-correlations can robustly
constrain the mean halo mass but cannot discriminate among distributions
with the same mean mass.

\subsection{Comparison to Linear Theory}
\label{sec:linth}

At sufficiently large scales, we expect the \lya\ flux auto-correlation and
the halo-flux cross-correlation to be adequately described by linear
perturbation theory. In redshift-space the linear theory 3-d flux power
spectrum is 
\begin{equation} 
P^F(k,\mu)=b_F^2(1+\beta_F\mu^2)^2P_m(k)~,
\end{equation} 
where $P_m(k)$ is the linear theory matter power spectrum,
$\mu$ is the cosine of the angle between the wavevector $\vec k$ and the
line of sight, and $b_F$ and $\beta_F$ are the bias factor and
redshift-space distortion parameter of the flux
\citep{Kaiser1987,McDonald2003}. For halos or galaxies one expects
$\beta=f(z)/b\approx[\Omega_m(z)]^{0.55}/b$, where $f(z)$ is the
fluctuation growth rate. However, for flux this relation does not hold
because of the non-linear relation between flux and optical depth, and
$\beta$ must be predicted separately (for more discussion see
\citealt{Seljak2012,Arinyo2015}). 
For the halo-flux cross power spectrum the linear
theory prediction is 
\begin{equation}
P^{Fh}(k,\mu)=b_Fb_h(1+\beta_F\mu^2)(1+\beta_h\mu^2)P_m(k)~, 
\label{eq:linpk}
\end{equation}
where the halo bias $b_h$ can be calibrated from numerical simulations
(T10) and $\beta_h=f(z)/b_h$. Correlation functions can be computed from
power spectra using the appropriate Fourier transform relations (e.g.,
equations $4.5-4.12$ of \citealt{Slosar2011}). The matter power spectrum
used for our linear theory calculations has the same WMAP7 parameters
adopted for the hydro and $N$-body simulations (see \S\ref{sec:sim}). We
measure $\xi(r,\mu)$ in bins of $\Delta\mu=0.05$ and compute the monopole
and quadrupole $\xi_0(r)$ and $\xi_2(r)$ from the sum
\begin{equation}
\xi_l(r) = (2l+1) \sum_{\mu=0}^{\mu=1} \xi(r,\mu) P_l(\mu) \Delta\mu~,
\end{equation}
where $P_0(\mu)=1$, $P_2(\mu)=\frac{3}{2}\mu^2-1$ are the
Legendre polynomials of degree $l$.

\begin{figure*}
\begin{minipage}{175mm}
\centering
\includegraphics[width=\linewidth]{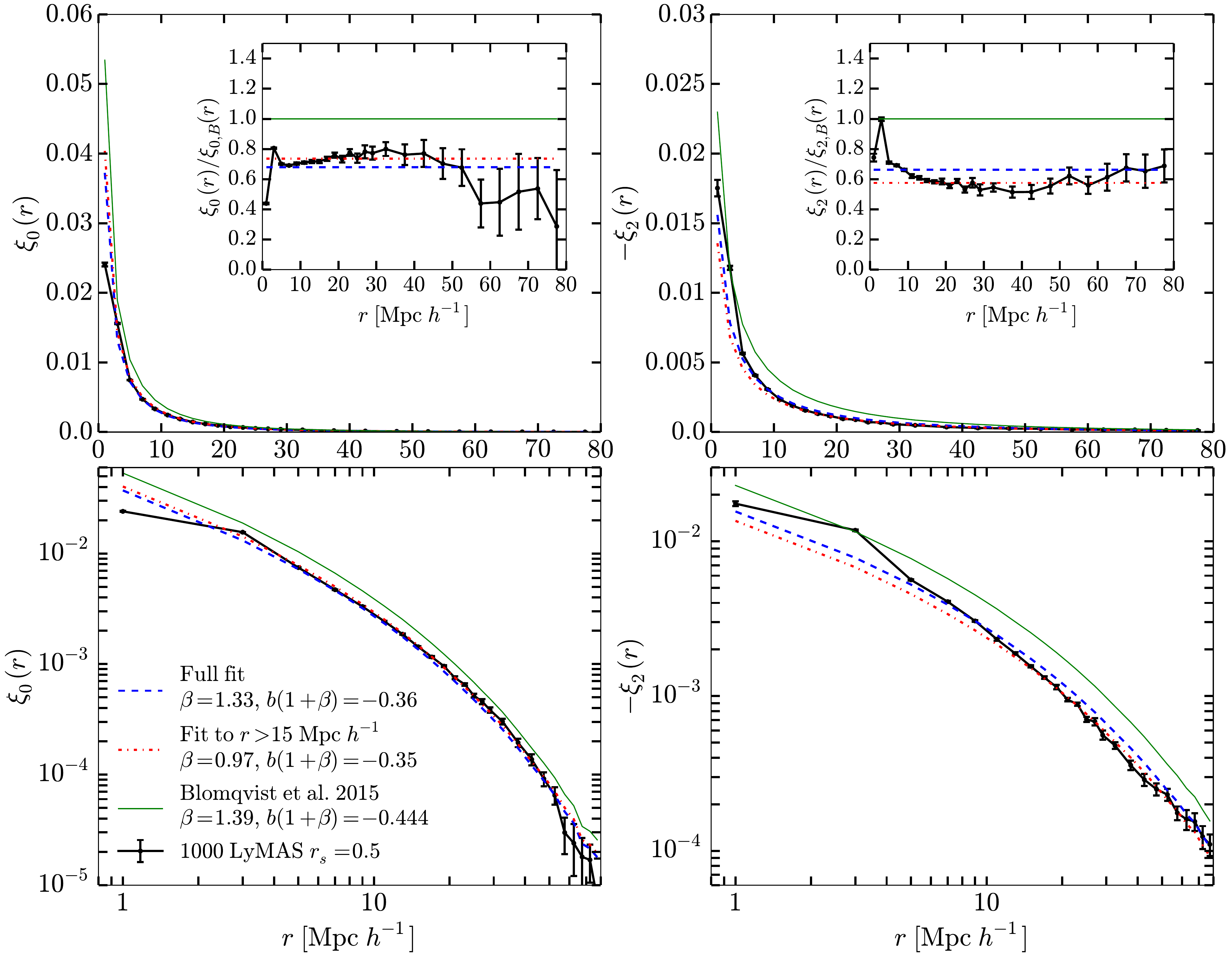}
\caption{The monopole (left) and quadrupole (right) of the auto-correlation
of \lya\ forest flux as determined from LyMAS in the (1 \hgpc)$^3$ volume
(solid black) compared to the linear theory model with parameters $b$ and
$\beta$ fit to the full range of data (blue dashed) and fit to only $r>15$
\hmpc\ (red dot-dashed). Green solid curves show linear theory with values
of $b$ and $\beta$ from the BOSS analysis of \citet{Blomqvist2015}, scaled
to $z=2.5$. Insets show ratio to linear theory with the
\citet{Blomqvist2015} values.}
\label{fig:auto}
\end{minipage}
\end{figure*}

Fig.~\ref{fig:auto} shows the monopole and quadrupole of the flux
auto-correlation in the (1 \hgpc)$^3$ LyMAS simulation as a function of
separation. Errors are the error in the mean among 16
subvolumes as described in \S\ref{sec:LvsH}. The top row shows the
correlation in linear space, and the bottom row is the same correlation in
log-log space. The dashed and dot-dashed lines in each panel show linear
theory predictions with $b_F$ and $\beta_F$ determined by fitting the full
range of data or only $r>15$ \hmpc, respectively.

The $\chi^2$ for the large separations is $43.23$ for 34 d.o.f. but the
fit to the full data range is extremely poor, with $\chi^2=4593$ for 48
d.o.f. Visually one can see that the fit to $r>15$ \hmpc\ gives a 
good description of the LyMAS monopole and quadrupole at these scales,
but the LyMAS quadrupole rises above this fit at smaller $r$.
At intermediate $r$ ($\approx 10-50$ \hmpc), the linear theory model fit to 
the full range underpredicts $\xi_0(r)$ and overpredicts $|\xi_2(r)|$. For 
$r>15$ \hmpc\ our best-fit parameters are $\beta=0.970\pm0.016$,
$b=-0.178$. There is significant degeneracy between these parameters, and
\lya\ forest analyses frequently quote the combination $b(1+\beta)$, which
is better determined than $b$ or $\beta$ alone. Our $r>15$ \hmpc\ fit
yields $b(1+\beta)=-0.3525\pm0.0011$.

Analyses of the BOSS \lya\ forest have reported a variety of values for
$\beta$ and $b(1+\beta)$ (e.g.,
\citealt{Slosar2011,Busca2013,Delubac2015}), with significant dependence on
details of the data analysis. The most precise, and likely most robust,
determination is from \citet{Blomqvist2015}, who implement a new method for
removing biases imprinted by the continuum determination procedure. They
find $\beta=1.39^{+0.11}_{-0.10}$ and $b(1+\beta)=-0.374\pm0.007$ at
$z=2.3$. They do not fit for redshift dependence, so here we scale the
value $b(1+\beta)$ by $[(1+2.5)/(1+2.3)]^{2.9}=1.19$ using the redshift
evolution reported by \citet{Slosar2011}. 
We have also multiplied the \citet{Blomqvist2015} predictions by a factor
$0.79/0.81$ to account for the slightly lower $\sigma_8$ value that
they assumed. With these parameters the
\citet{Blomqvist2015} fit lies above the LyMAS prediction, by factors of
approximately 1.25 and 1.6 for the monopole and quadrupole, respectively.
If we did not apply the redshift scaling to $b(1+\beta)$ these factors
would be approximately $0.95$ and $1.25$.

\begin{figure*}
\begin{minipage}{175mm}
\centering
\includegraphics[width=\linewidth]{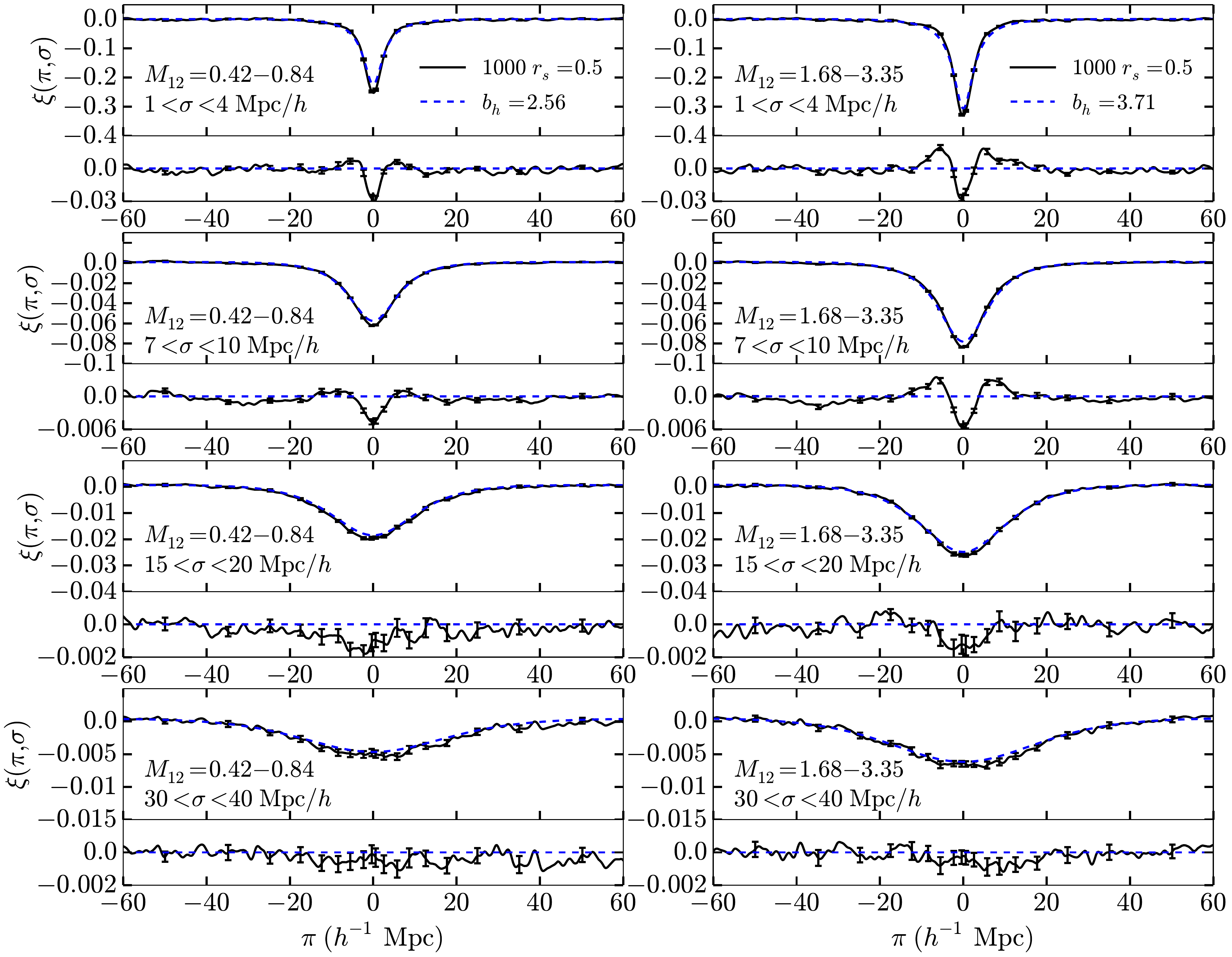}
\caption{The cross-correlation for four transverse separation bins,
$\sigma=1-4,\ 7-10,\ 15-20,\ 30-40$ (rows) and two halo mass bins \Mtwel\
$=0.42-0.84,\ 1.68-3.35$ (columns) in the (1 \hgpc)$^3$ LyMAS box (black
solid) compared to the cross-correlation predicted by the linear theory for
redshift distortion $\beta=0.97$, \lya\ forest bias
$b_{\mathrm{Ly}\alpha}=-0.178$, and the T10 value for halo bias in these
mass bins, $b_h=2.56$ or $b_h=3.71$ (blue dashed).}
\label{fig:cross_theory}
\end{minipage}
\end{figure*}

We leave investigation of this discrepancy in the auto-correlation 
function to future work. There are a number of possible contributions, 
and it is not clear whether one dominates or a variety of small effects 
combine. One is incorrect cosmological parameters in the Horizon-noAGN 
simulation, though the WMAP7 parameters used here are only moderately different 
(most notably in $\Omega_m$) from the Planck 2015 values \citep{Planck2015}.
A second contribution could be inaccurate IGM properties: our adopted value of 
$\bar{F}=0.795$ (based on \citealt{Faucher-Giguere2008}) could be incorrect, or
the UV background history used in Horizon-noAGN could lead to an incorrect 
temperature-density relation in the diffuse IGM, perhaps because it does not 
include extra heating by helium reionization. A related possibility, and one 
that might be more capable of producing substantial differences in $b$ and 
$\beta$, is that spatial variations in the UV background or the IGM 
temperature-density relation induce additional large scale clustering of the 
\lya\ forest flux that is not accounted for in our simulations \citep[see 
recent discussions by][]{Pontzen2014,Gontcho2014,Greig2015}. Another possible 
explanation is that LyMAS does not reproduce full hydro simulation results for 
the flux auto-correlation with sufficient accuracy. Although we find excellent 
agreement between hydro and LyMAS calculations of the halo-flux 
cross-correlation in the (100 \hmpc)$^3$ simulations 
(Fig.~\ref{fig:LyMAS_vs_hydro}), the auto-correlation monopole and quadrupole 
differ at the $10-20\%$ level at $r\sim10$ \hmpc, while differences at larger 
scales are harder to assess because the correlation function is artificially 
suppressed by the finite box size (W. Zhu, private communication). The mismatch 
in Fig.~\ref{fig:auto} is cause for some caution about the mass or bias values 
that we infer from fitting BOSS cross-correlation measurements in \S\ref{sec:obs}.
However, we also note that different methods for inferring $\beta$ and $b$ from 
BOSS have yielded a wide range of values because of the challenges of correcting 
for continuum determination biases, metal-line contamination, and DLA 
absorption, and the \citet{Blomqvist2015} values may not be the last word. There 
is also uncertainty in scaling from $z=2.3$ to our simulation redshift of 
$z=2.5$.

Recently, \citet{Arinyo2015}, extending the work of \citet{McDonald2003}, 
have attempted to compute $b$ and $\beta$ theoretically using 3-d flux power 
spectra measured from a number of hydro simulations with comoving box sizes 
ranging from $40-120$ \hmpc. Because of the limited box size, these authors 
fit non-linear models to the numerical $P_F(k,\mu)$, and the extrapolated large 
scale values of $b$ and $\beta$ depend on the assumed non-linear models as well
as on simulation and cosmological parameters. Their estimates of $\beta$ at 
$z\approx 2.5$ are typically $1.2-1.4$, clearly higher than the value we find 
here from LyMAS. The values of $b$ are harder to compare because they are more 
sensitive to cosmological parameters and the adopted form of the non-linear 
model. LyMAS and the model-based extrapolation method of \cite{Arinyo2015}
represent two different strategies for deriving large scale \lya\ forest 
clustering from limited volume hydro simulations, and over time we hope that 
they will converge to yield consistent predictions given the same IGM and 
cosmological parameters.

\begin{figure*}
\begin{minipage}{175mm}
\centering
\includegraphics[width=\linewidth]{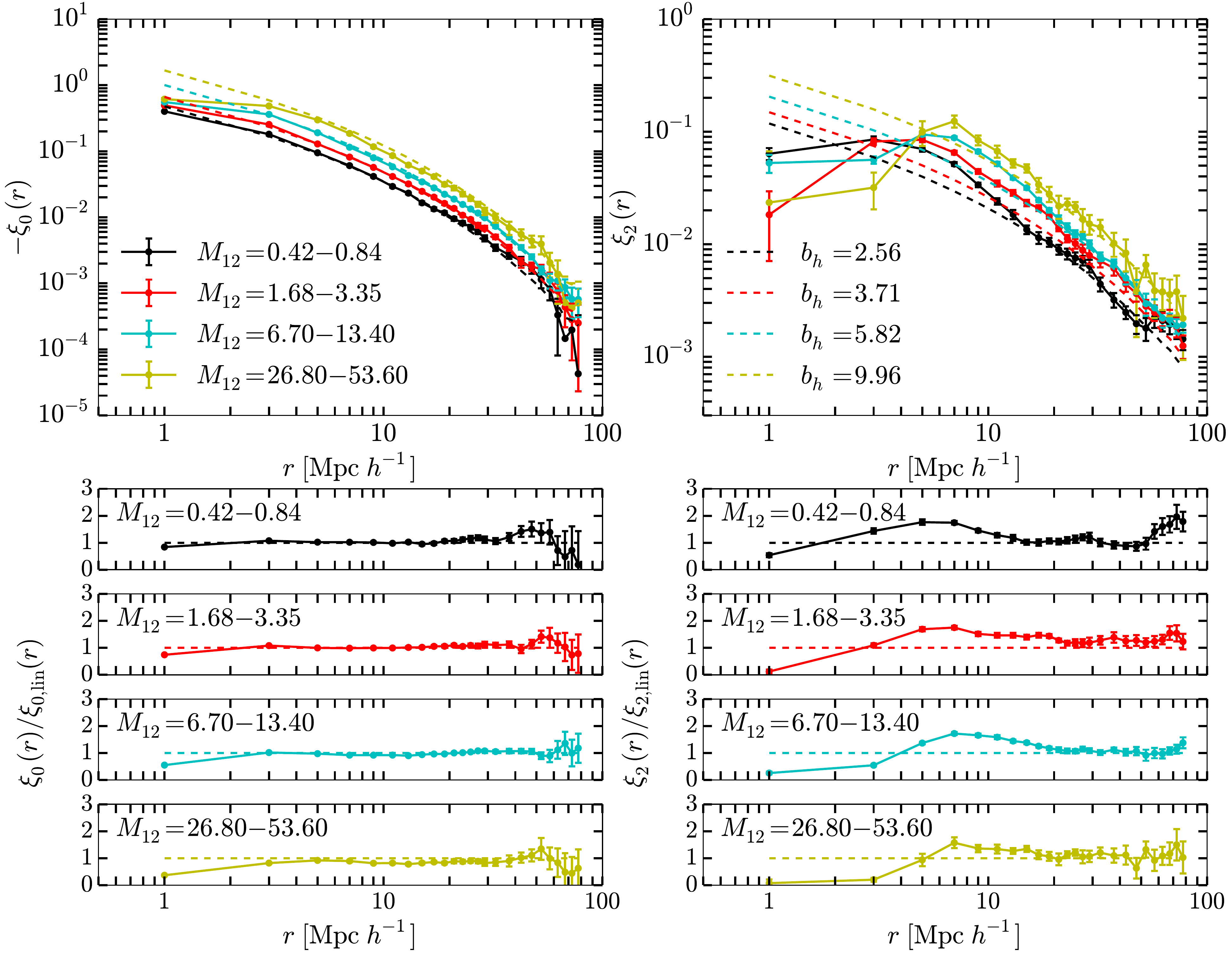}
\caption{The monopole (left) and quadrupole (right) of the
cross-correlation between halo centers and the \lya\ forest. 
LyMAS results for four mass bins, \Mtwel\
$=0.42-0.84,\ 1.68-3.35,\ 6.70-13.40,$ and $26.80-53.60$, are shown as the
solid lines connecting points with errors (mass increases bottom to top).
Dashed lines show the linear theory calculation for the $b_F$ and $\beta_F$
determined by fitting the LyMAS flux auto-correlation and 
halo biases $b_h$ that correspond to
the log-center of each mass bin shown using the \citet{Tinker2010} $b_h(M)$
relation, with halo redshift distortion $\beta_h=1/b_h$ (bias increases bottom to top). Bottom
panels show ratio of LyMAS cross-correlation to the linear theory
calculation.}
\label{fig:cross_monoquad}
\end{minipage}
\end{figure*}

Returning to the cross-correlation that is the focus of this paper, 
Fig.~\ref{fig:cross_theory} compares the halo-flux \xips\ from LyMAS to the
linear theory model for two different halo mass ranges. In this case, the
linear theory is that for the cross-correlation between \lya\ forest and
halos, with redshift distortion parameter $\beta_F=0.97$ and \lya\ bias
$b_F=-0.178$, equivalent to our best-fit parameters to the
$r>15$ \hmpc\ fit to the LyMAS auto-correlation. For the \Mtwel\
$=0.42-0.84$ halo mass range, the T10 bias is $b_h=2.56$, and the T10 halo
bias for the larger \Mtwel\ $=1.68-3.35$ bin is $b_h=3.71$. With these
bias values the LyMAS \xips\ visually matches linear theory closely even down
to $\sigma = 1-4$ \hmpc\ and $\pi \approx 0$.
However, there are statistically significant residuals for small $\pi$ at
$\sigma=1-4\ \mhmpc$ and $7-10\mhmpc$, with peak differences
about 10\% of $\xi(\pi=0,\sigma)$.

\begin{figure}
\centering
\includegraphics[width=\linewidth]{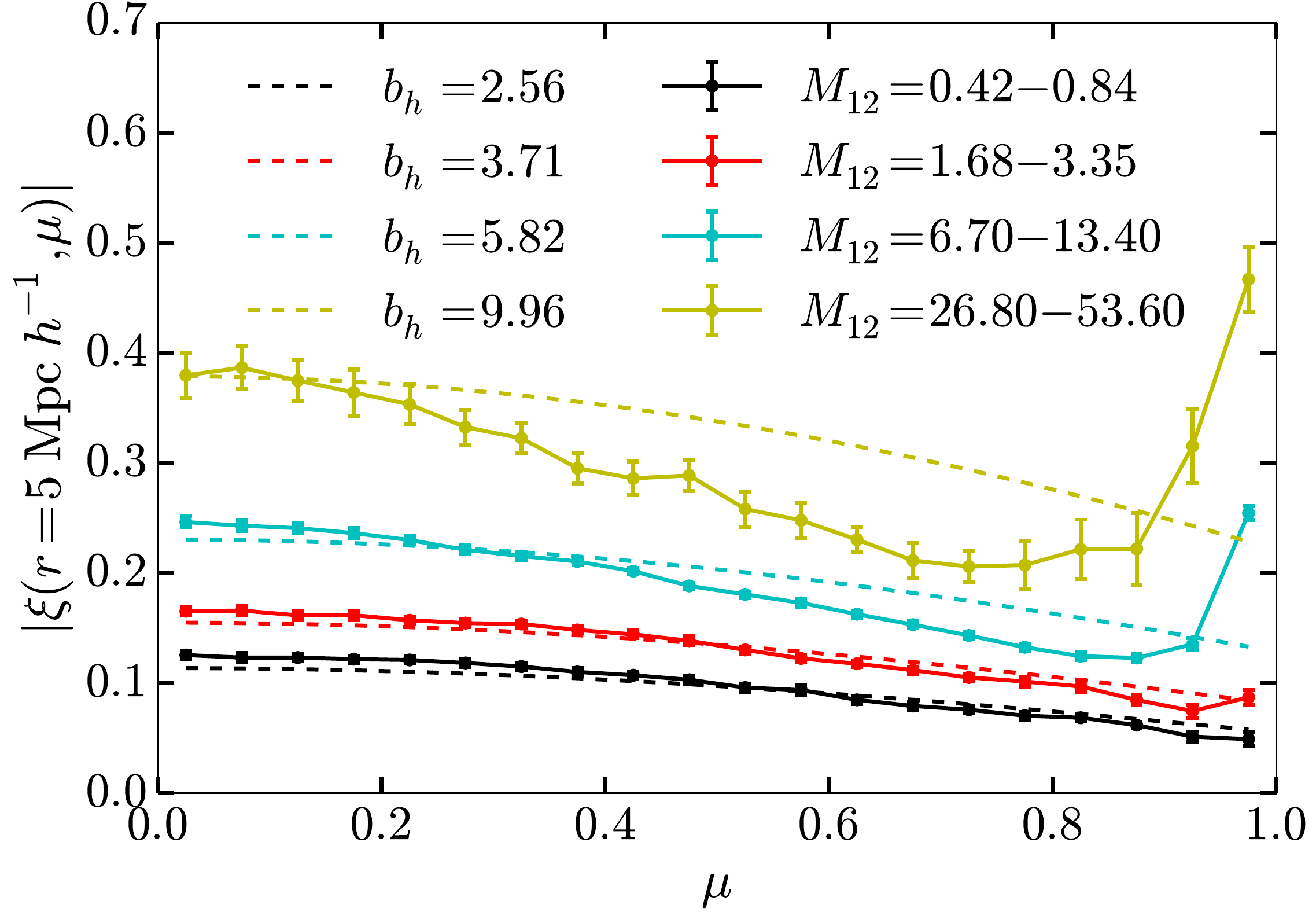}
\caption{The cross-correlation between halo centers and the \lya\ forest
from LyMAS at a separation of $r=5$ \hmpc\ as a function of the cosine of the
angle from the line of sight, $\mu$, for the halo mass bins \Mtwel\
$=0.42-0.84,\ 1.68-3.35,\ 6.70-13.40,$ and $26.80-53.60$ (mass increases bottom to top). Solid lines show
the correlation predicted by LyMAS and dashed lines show the linear theory
prediction (bias increases bottom to top).}
\label{fig:cross_r5}
\end{figure}

Fig.~\ref{fig:cross_monoquad} examines the comparison between LyMAS and
linear theory in terms of the monopole and quadrupole of the halo-flux
cross-correlation. We use the same $\beta_F$ and $b_F$ values
as in Fig.~\ref{fig:cross_theory} and the halo bias factors from the T10
formula evaluated at the logarithmic center of each mass bin. The monopole 
is well described by linear theory at
$r>3$ \hmpc, with some deviation for the highest mass bin \Mtwel\ $=
26.8-53.6$. However, the quadrupole rises above linear theory at $r=10-20$
\hmpc, then flattens below it at $r < 5$ \hmpc. Thus, the level of
agreement between LyMAS and linear theory in Fig.~\ref{fig:cross_theory} is
to some degree a consequence of the particular quantity plotted. 
Fig.~\ref{fig:cross_r5} plots $|\xi(\mu)|$ for comoving separation $r=5$ \hmpc. 
The two lowest mass bins show reasonably good accord with linear theory at all 
$\mu$, with LyMAS rising slightly above for separations perpendicular to the line of 
sight ($\mu\approx 0$) and falling slightly below for separations along the 
line of sight ($\mu \approx 1$). However, for the two highest mass
bins LyMAS yields substantially weaker correlations at $\mu > 0.5$
(\Mtwel\ $=6.7-13.4$) or $\mu > 0.3$ (\Mtwel\ $=26.8-53.6$). The overall
normalization of the linear theory curves, and to a lesser degree the $\mu$
dependence, depends on the $b_F$ and $\beta_F$ values inferred from the
LyMAS flux auto-correlation, and the volume of the (100 \hmpc)$^3$ hydro
simulation limits our ability to test the accuracy of LyMAS in predicting
these quantities.

In summary, for halo masses in the range \Mtwel\ $\approx 0.5-3$ found by
FR12 and FR13, LyMAS predictions for \xips\ in bins of $\sigma$ follow
linear theory even down to small scales. However,
other representations show significant deviations between LyMAS and linear
theory, especially at higher halo masses and high values of $\mu$.

\section{Comparison to Observed Cross-Correlations}
\label{sec:obs}

Now that we have examined how the cross-correlation depends on dark matter
halo mass and separation from the halo (in both line-of-sight and
transverse directions), we can compare the halo-forest correlation to other
cross-correlations with the forest, such as quasars and DLAs. This will
allow us to determine a characteristic mass for the host halos of these
objects.

\begin{figure*}
\begin{minipage}{175mm}
\centering
\includegraphics[width=\linewidth]{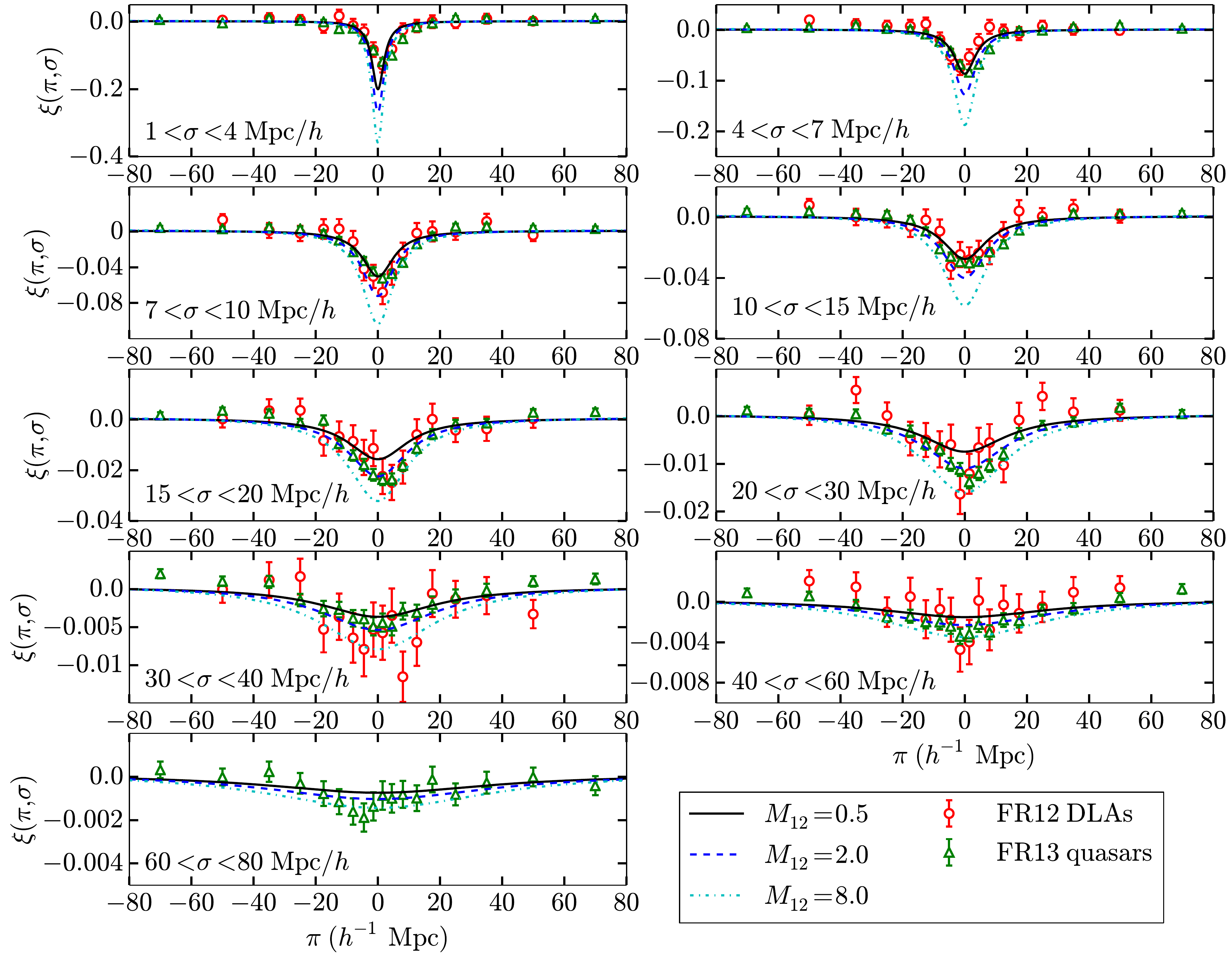}
\caption{The cross-correlation as predicted by LyMAS (described by
Lorentz profile fits, and adding the mean transmission correction) 
for halo masses of \Mtwel\ $=0.5$ (black solid),
\Mtwel\ $=2.0$ (blue dashed) and \Mtwel\ $=8.0$ (cyan dot-dashed). Red circles and green
triangles are the cross-correlation measured by FR12 and FR13 for
DLAs and quasars, respectively. Panels show
different transverse separation bins as indicated, increasing top to bottom, 
left to right.}
\label{fig:fitted_FR}
\end{minipage}
\end{figure*}

The measured cross-correlations of the \lya\ forest with DLAs and with
quasars (FR12, FR13) are plotted in Fig.~\ref{fig:fitted_FR} as the red circles and
green triangles, respectively. These correlations were measured using $\sim$ 60,000 quasar
spectra in DR9 of BOSS; measurements from the final, DR12 BOSS sample are
in progress. A surprising feature of the data points is that the DLA and
quasar cross-correlations are similar in amplitude, even though FR12 and
FR13 derive quite different bias factors ($\bdla=2.17\pm 0.2$, $b_Q=3.64\pm
0.13$) for these two populations. However, the FR13 $b_Q$ fit is driven by
the large $r=\sqrt{\pi^2+\sigma^2}$ bins, and the $r<15$ \hmpc\ bins are
discarded, while smaller scale measurements carry much of the weight in the
FR12 $\bdla$ fits. We have calculated the correlation function in LyMAS
in the same transverse separation bins as the measured correlation to
obtain the closest comparison. Lorentz profiles with amplitude, width, and
offset predicted by our LyMAS calculation for halos of mass \Mtwel\ $=0.5$
(black solid), 2.0 (blue dashed), and 8.0 (cyan dot-dashed), are determined by the 
equations~(\ref{eq:depth}) and~(\ref{eq:FWHM}) and Tables~\ref{tab:mass_params} 
and~\ref{tab:delta}.

Because BOSS \lya\ forest spectra are calibrated to reproduce the mean
\lya\ absorption as a function of redshift, there is a slight bias in
cross-correlation measurements, since the positive correlation over the
full \lya\ forest is calibrated out. FR12 and FR13 apply a Mean
Transmission Correction (MTC) that removes this bias. The derivation of
the MTC is laid out in Appendix A of FR12. We apply this correction to our
predicted cross-correlation whenever we compare to the measured correlation
of quasars or DLAs. Generally, the correction shifts correlations upward
so that the cross-correlation at large $|\pi|$ becomes positive instead of
near zero. It is a small effect, but important for fitting the non-zero
values of the measured cross-correlation.

In the $\sigma\approx 10-40$ \hmpc\ transverse separation bins, it appears
as though the cross-correlation for halos of average mass \Mtwel\ $=0.5$ or
2.0 fits the measurement of the cross-correlation for DLAs or quasars
fairly well. However, the correlation for halos of the same mass in the
$\sigma=1-4$ \hmpc\ separation bin does not fit the data. We expect that,
if there is a characteristic mass distribution for halos that host DLAs or
quasars, then the correlation predicted for that mass distribution should
match the observations for all separation bins. The drastic overestimation
of the strength of the correlation in the smaller transverse separation
bins indicates a missing component in our model of the halo-flux
cross-correlation.

\begin{figure}
\centering
\includegraphics[width=\linewidth]{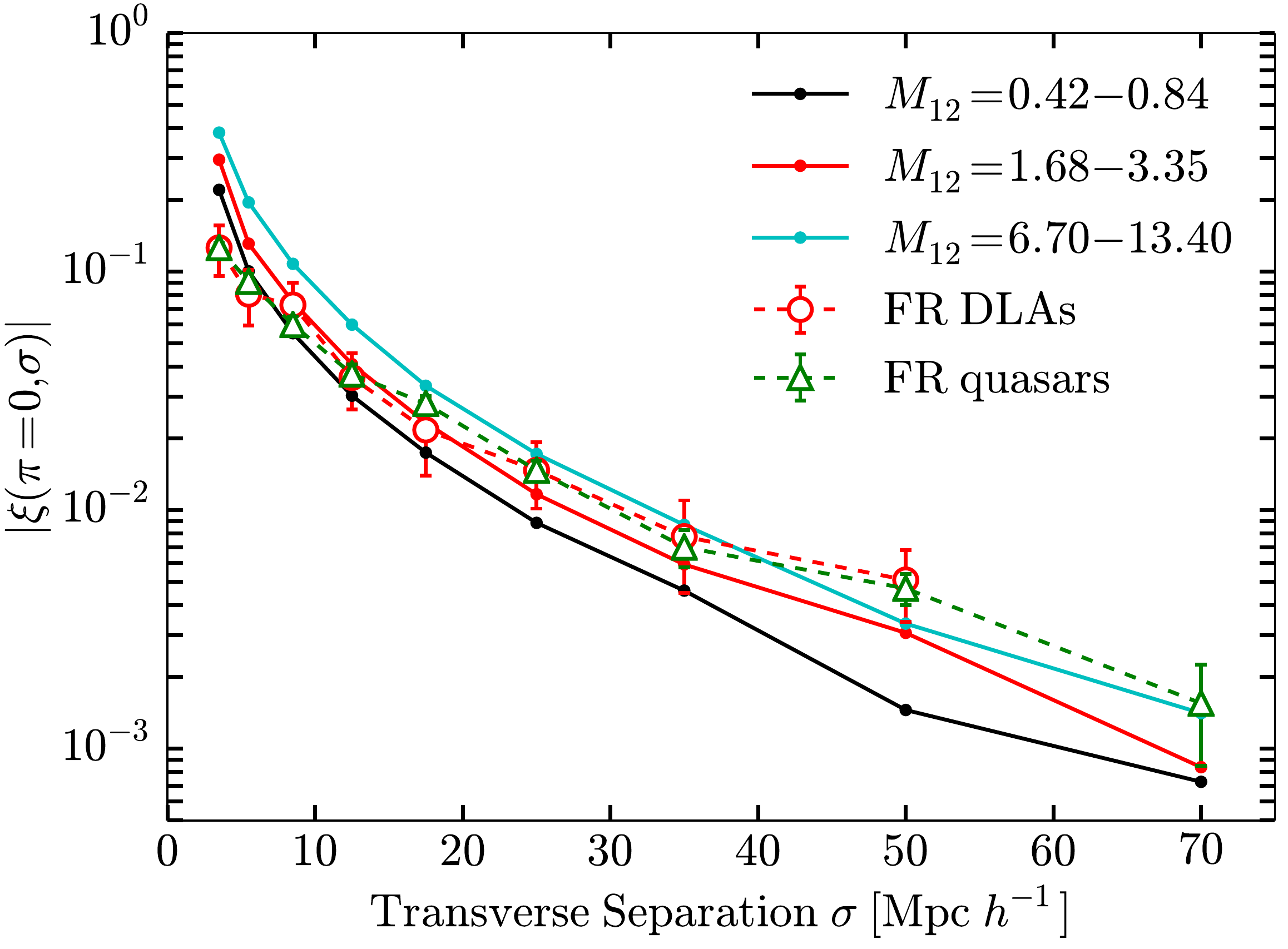}
\caption{The strength of the cross-correlation as measured by the best-fit
amplitude of the Lorentz profile fit to LyMAS for each of the different
halo mass bins (solid lines; mass increases bottom to top) and transverse separation bins. The
red dashed and green dashed lines with open points show the amplitude of
the measured correlation for DLAs and quasars, respectively, found by
fitting the Lorentz profile to the data.}
\label{fig:depvstrsep}
\end{figure}

Fig.~\ref{fig:depvstrsep} summarizes this data-model comparison by fitting
our Lorentz profile form to the DLA and quasar data in each $\sigma$ bin.
We show the measured amplitude as a function of $\sigma$ by circles and
triangles for DLAs and quasars, respectively, while solid curves show the
amplitude of Lorentz profiles fit to the simulation cross-correlations for
three different halo mass ranges. As in Fig.~\ref{fig:fitted_FR}, it is
evident that the DLAs and quasars exhibit similar overall clustering
strength but that any halo mass that matches the observed clustering
strength at $\sigma=20-50$ \hmpc\ overpredicts it at $\sigma=1-10$ \hmpc.
If our predictions were based on linear theory one might conclude that this
trend reflects the breakdown of linear theory predictions, but LyMAS is a
fully non-linear method and we have shown (Fig.~\ref{fig:LyMAS_vs_hydro})
that it reproduces hydrodynamic simulation results essentially perfectly even at
$\sigma=1-4$ \hmpc. The discrepancy could reflect some form of feedback or
IGM physics not incorporated into our hydro simulations, but the complete
insensitivity of the halo-flux predictions to the presence or absence of
AGN feedback (Fig.~\ref{fig:AGN}) makes such an explanation unlikely.

One key difference between the simulations and the observations is that we
know the redshift-space position of simulated halos exactly while the
redshifts of quasars and DLAs must be estimated observationally from
emission and absorption lines. Internal motions and radiative transfer
effects lead to differences of many hundreds of km s$^{-1}$ among quasar
redshifts estimated from different lines
\citep{Espey1989,Richards2002,Hewett2010}. From the asymmetry of \xips,
FR13 concluded that the quasar redshifts reported by the BOSS pipeline had
a systematic offset of $-157$ km s$^{-1}$, which is small but not
completely negligible compared to the width of \xips\ in bins of small
$\sigma$. More importantly, random redshift errors with a velocity
distribution $f(\pi)$ will convolve the predicted \xips\ in the $\pi$
direction, increasing the width but decreasing the depth of
\xips.\footnote{This statement holds true even though our $\sigma$ bins
have a finite width, a point we have tested explicitly by adding random
offsets to halo positions and remeasuring \xips. Note that an offset of
$\pi$ comoving \hmpc\ corresponds to a velocity offset
$v=\pi\times(1+z)^{-1}\times(100\ \mathrm{km\ s}^{-1})\times H(z)/H_0$ at
redshift $z$. For our cosmology at $z=2.5$, $v=101\pi\ \mathrm{km\ s}^{-1}$.}
This convolution has a significant impact when $\sigma$ is
comparable to the RMS redshift error but becomes negligible at large
$\sigma$, so it has the correct qualitative form to reconcile our
predictions with the BOSS measurements.

\begin{figure*}
\begin{minipage}{175mm}
\centering
\includegraphics[width=\linewidth]{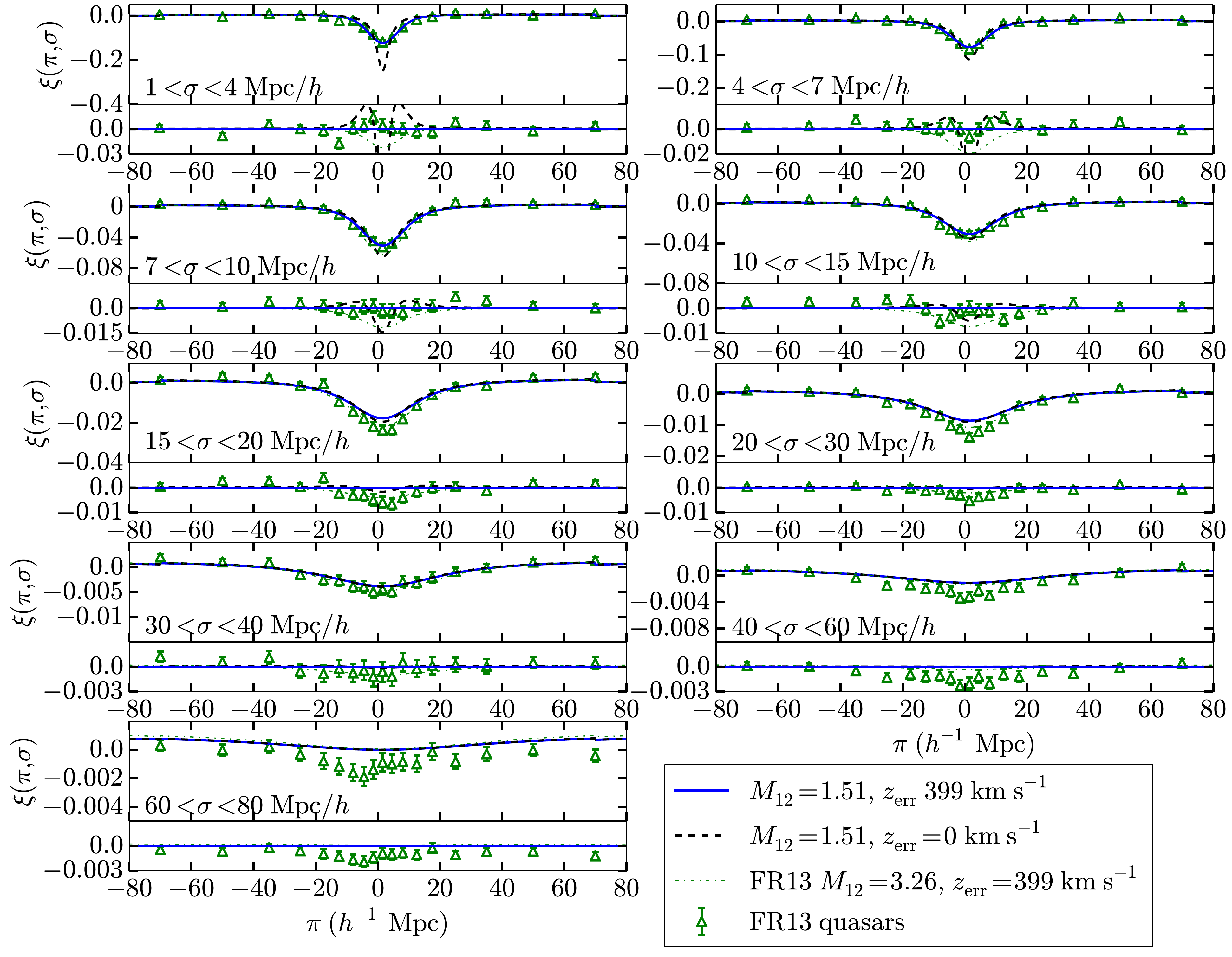}
\caption{The Lorentz profile that describes the cross-correlation for a
halo mass of \Mtwel\ $=1.51$ (black dashed) and this same correlation when
a redshift measurement error of 399 km s$^{-1}$ (blue solid) is included in
the line-of-sight position of the dark matter halos. This is the error
that brings the strength of the correlation closest to that of the data.
Green triangles are the cross-correlation between quasars and the \lya\
forest from \citet{FR2013}, and green dot-dashed line is the correlation implied
by FR13's best-fit bias. Panels show different transverse separation bins
as labeled. Lower panels show residuals from the best-fit model with 
$z_{\mathrm{err}}=399$ km s$^{-1}$.}
\label{fig:QSO_zerr}
\end{minipage}
\end{figure*}

Fig.~\ref{fig:QSO_zerr} presents a fit to the FR13 data in which we add
their inferred mean $\pi$ offset of 157 km s$^{-1}$ to the model
predictions and add the RMS velocity error as a free parameter along with
halo mass. We use the full covariance matrix that FR13 estimated analytically 
by treating the forest as a Gaussian random field.
There are significant off-diagonal
correlations of errors at different $\pi$ in a given $\sigma$ bin, but
correlations across $\sigma$ bins are generally weak. In our fit we adopt
the Lorentz function form of equation~(\ref{eq:Lorentz}) convolved with a
Gaussian distribution of redshift errors, using equation~(\ref{eq:FWHM})
for the Lorentzian width, the power-law form of equation~(\ref{eq:depth})
to predict amplitude as a function of halo mass (with parameters listed in
Table~\ref{tab:mass_params}), and the best-fit offset $\Delta$ as a
function of halo mass (determined by interpolating the offsets given in
Table~\ref{tab:delta}). When inferring errors on the halo mass \Mtwel\ and
RMS velocity error we incorporate only the observational measurement
uncertainties, not the uncertainties in the parameters describing our fits
to the LyMAS simulation, which would have a small relative effect.

It is evident from Fig.~\ref{fig:QSO_zerr} that including the RMS redshift
error as a parameter dramatically improves the fit to the FR13
measurements, eliminating the systematic mismatch between large and small
$\sigma$ bins seen in Figs.~\ref{fig:fitted_FR} and~\ref{fig:depvstrsep}.
The best-fit RMS error is $399\pm21$ km s$^{-1}$, corresponding to
$\pi=3.8$ \hmpc\ for our WMAP7 cosmological parameters. 
This RMS error appears plausible relative to comparisons of 
different redshift estimators as described in the BOSS DR9
quasar catalog paper of \cite{Paris2012}.
The global
best-fit halo mass is \Mtwel\ $=1.51^{+0.11}_{-0.10}$. As discussed in
\S\ref{sec:Mh}, this should be interpreted as a bias-weighted mean host
halo mass of quasars, i.e., it is the mass $\bar M$ for which
\begin{equation}
b(\bar M)=\frac{\int_0^\infty \mathrm{d}M\
b(M)p(M)\mathrm{d}n/\mathrm{d}M}{\int_0^\infty \mathrm{d}M\
p(M)\mathrm{d}n/\mathrm{d}M}
\end{equation}
where $p(M)$ is the probability
that a halo of mass $M$ hosts a BOSS quasar at any given time and d$n$/d$M$
is the halo mass function. With the T10 halo bias relation at $z=2.5$, 
our halo mass range corresponds to $b(\bar M)= 3.18\pm 0.06$.
Evaluated at the central redshift of the observational measurement,
$z=2.3$, the same halo mass corresponds to $b(\bar M)=2.92\pm0.06$. 

One can see from Fig.~\ref{fig:QSO_zerr} that the LyMAS predictions
including RMS velocity errors are a good but not perfect fit to the FR13
measurements, given the extremely small statistical errors of the latter.
The overall $\chi^2$ is 290 for 162 data points, with two fitting
parameters. Fig.~\ref{fig:cosvar} also shows residuals between the LyMAS
numerical results and our Lorentzian fits to those results, but those
residuals are small compared to the differences between the model
predictions and the data. We have checked that adding these residuals
(scaled in amplitude by the halo bias) to our convolved Lorenzian model ---
in effect, using the direct simulation results in place of the Lorentzian
description of them --- does not noticeably change the $\chi^2$ of the fit
or the best-fit values of the model parameters.

\begin{figure*}
\begin{minipage}{175mm}
\centering
\includegraphics[width=\linewidth]{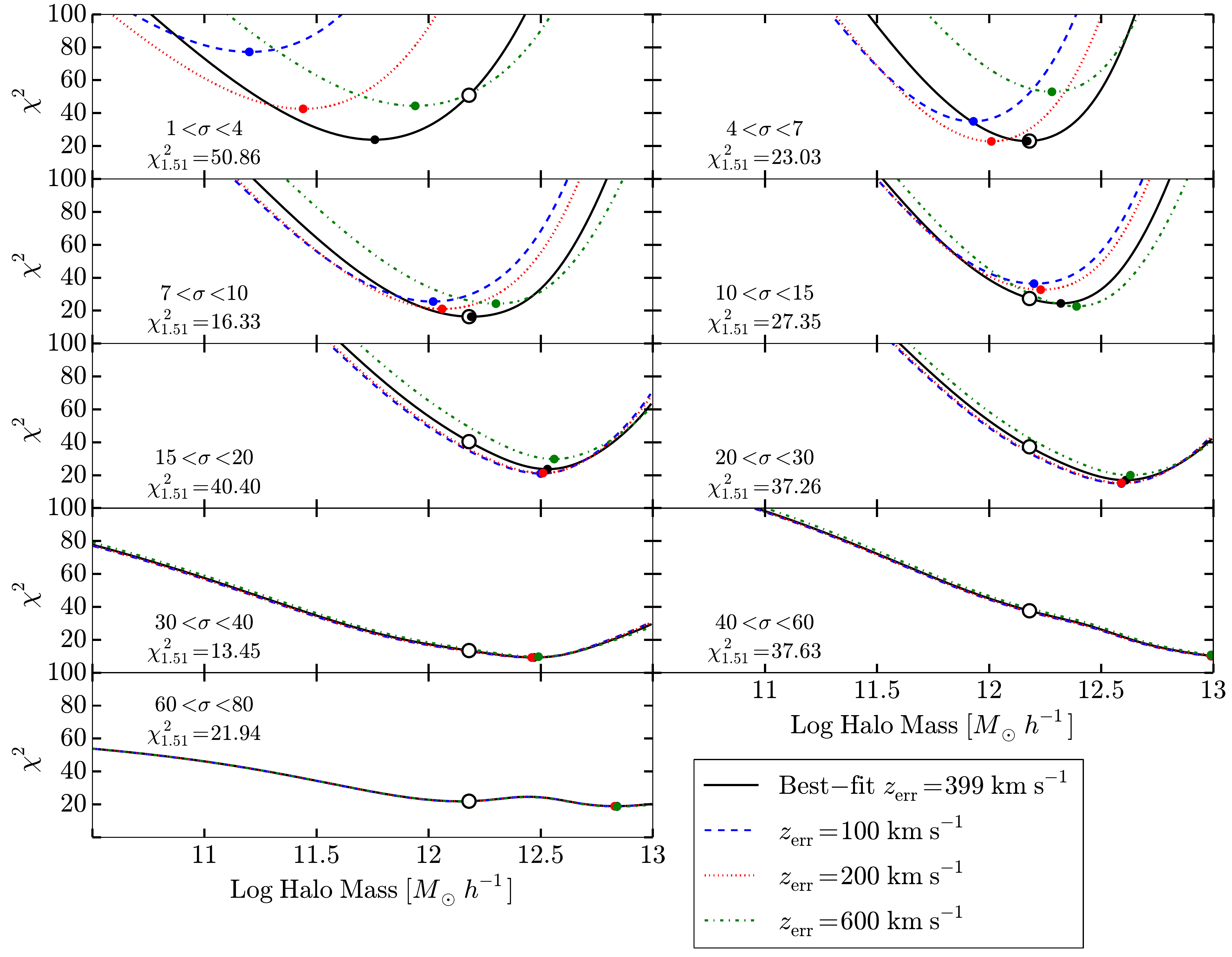}
\caption{$\chi^2$ as a function of halo mass in each $\sigma$ bin of the
FR13 quasar-flux cross-correlation measurement. Black curves adopt our
global best-fit value of 399 km s$^{-1}$ for the RMS velocity error,
while dashed, dotted, and dot-dashed curves impose RMS velocity errors of 100, 200,
and 600 km s$^{-1}$, respectively. Filled circles mark the minimum of each
curve, i.e. the halo mass preferred by the data in this $\sigma$ bin alone
for the adopted velocity error. Open black circles show the global
best-fit \Mtwel\ $=1.51$, and the contribution to $\chi^2$ from each
$\sigma$ bin for this best-fit model is listed. There are 18 \xips\ data
points in each $\sigma$ bin.}
\label{fig:QSO_chisq}
\end{minipage}
\end{figure*}

Fig.~\ref{fig:QSO_chisq} examines the $\chi^2$ contributions in greater
detail, with each panel showing the $\chi^2$ from data points in a single
$\sigma$ bin as a function of $M_{12}$, for our best-fit RMS velocity error
of $399\,\kms$ and for velocity errors of 100, 200, or $600\ \kms$. 
With a small velocity error, the
$\sigma=1-4\ \mhmpc$ bin has a high $\chi^2$ for any $M_{12}$, and, as seen
previously in Fig.~\ref{fig:fitted_FR}, the value of $M_{12}$ that would
match in this bin is strongly discordant with the values that fit larger
$\sigma$ bins. Conversely, a large velocity error makes $\xi(\pi,\sigma)$
too broad in the innermost bins, worsening $\chi^2$. Even with the
best-fit RMS error, the innermost bin favors a lower $M_{12}$ than other
bins. Apart from this bin, the $\sigma$ bins with the largest $\chi^2$
contributions are $15-20\ \mhmpc$, $20-30\ \mhmpc$, and $40-60\ \mhmpc$, where
one can see visually in Fig.~\ref{fig:QSO_zerr} that our best-fit
model underpredicts the measured correlation. In other bins the 
global best-fit $M_{12}=1.51$ yields close to the minimum $\chi^2$
contribution, and the $\chi^2$ is not very far from the
value $18\pm 4.2$ expected for 18 data points.

The slight mismatch between our simulation output at $z=2.5$
and the central redshift $z=2.3$ of the FR13 observational analysis
introduces some ambiguity in comparing our inferred halo bias and
halo mass to the results of FR13's linear theory fit.
(Ongoing work with the much larger BOSS DR12 sample will allow
tailored comparisons in fairly narrow redshift bins.)
Our value of $b(\bar M) = 3.18 \pm 0.06$ is formally 
inconsistent with the $b_Q = 3.64^{+0.13}_{-0.15}$ found by FR13,
and our inferred halo mass is correspondingly lower.
Using the T10 halo formula at $z=2.3$, the FR13 best-fit bias
corresponds to $M_{12}=3.26$. 
Green curves in Fig.~\ref{fig:QSO_zerr} 
show LyMAS predictions for this halo mass at $z=2.5$, with our best-fit
RMS velocity error. While this comparison is inexact because
of the redshift mismatch, it appears likely that most of the
difference between our inferred bias and the somewhat higher value
of FR13 arises because FR13 fit to points with $r > 15\ \mhmpc$,
thus matching the $\sigma = 15-60\ \mhmpc$ bins well and not
penalizing the overly strong correlations predicted at small $\pi$
in the smaller $\sigma$ bins.
The poor global $\chi^2$ of our fit indicates that
one should be cautious about our formal error bar. Furthermore, it is
clear from Fig.~\ref{fig:QSO_chisq} that higher $M_{12}$ is disfavored
mainly by the innermost $\sigma$ bin and is therefore sensitive to our
assumption of Gaussian velocity errors. If we keep the best-fit velocity
error but drop the innermost bin from the mass determination we get $M_{12}
= 2.19^{+0.16}_{-0.15}$, with corresponding $b(\bar M) = 3.63^{+0.08}_{-0.07}$
at $z=2.5$.

The mild tension between the halo mass and bias factor inferred
here and in FR13 is an indication of the current level of systematic
uncertainty associated with model fitting and data selection.
The FR13 value has the virtue of using a model that has an acceptable
$\chi^2$ over the range fitted ($r>15\ \mhmpc$) and matching the
central redshift of the data. It relies on the value of $b_F(1+\beta_F)$
inferred from fitting the observed flux auto-correlation, and this
fitting has statistical and systematic uncertainties of its own.
LyMAS directly predicts the halo-flux cross-correlation and thus
circumvents this step, but the mismatch between the LyMAS flux auto-correlation
and the \cite{Blomqvist2015} linear theory parameters 
(Fig.~\ref{fig:auto}) could indicate that the IGM structure
in the hydrodynamic simulations on which LyMAS is calibrated are incorrect. 
The excellent match between LyMAS and the full hydrodynamic simulation 
predictions of the cross-correlation indicates this is not an issue with 
LyMAS itself, but it could be a consequence of the physics and parameters 
assumed in the hydrodynamic simulations. The principal strength of the 
LyMAS mass determination is its use of a fully non-linear model that describes 
$\xi(\pi,\sigma)$ down to small scales. A hydrodynamic simulation on its own 
cannot make this prediction accurately because of limited volume, but the 
combination of hydrodynamics and LyMAS can.

\begin{figure*}
\begin{minipage}{175mm}
\centering
\includegraphics[width=\linewidth]{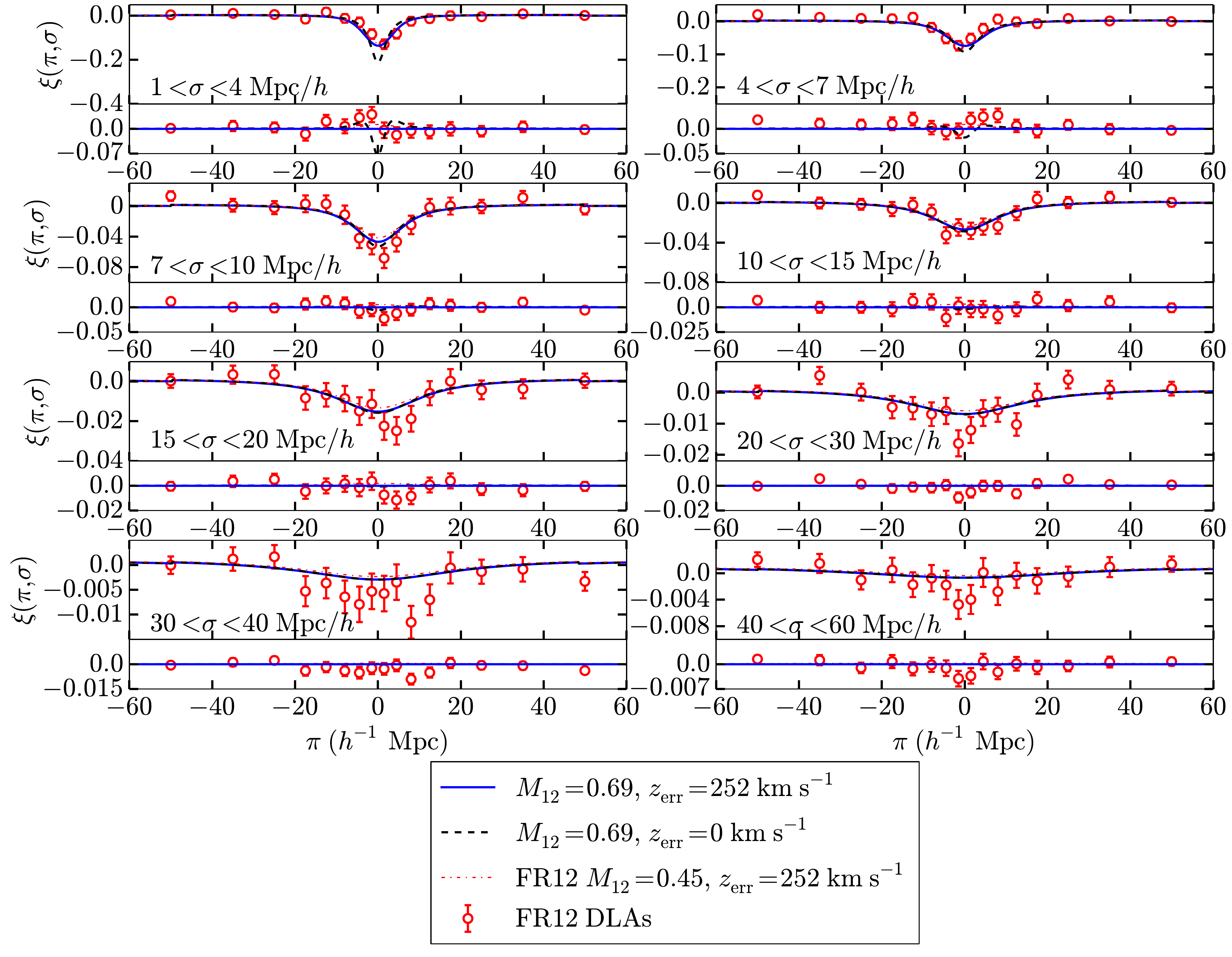}
\caption{The Lorentz profile that describes the cross-correlation for a
halo mass of \Mtwel\ $=0.69$ (black dashed) and this same correlation when
a redshift measurement error of 252 km s$^{-1}$ (blue solid) is included in
the line-of-sight position of the dark matter halos. This is the error
that brings the strength of the correlation closest to that of the data.
Red circles are the cross-correlation between DLAs and the \lya\ forest
from \citet{FR2012}, and red thin line is the correlation implied by FR12's
best-fit bias. Panels show different transverse separation bins as labeled. 
Lower panels show residuals from the best-fit model with $z_\mathrm{err}=252$ km s$^{-1}$.}
\label{fig:DLA_zerr}
\end{minipage}
\end{figure*}

Fig.~\ref{fig:DLA_zerr} presents our fit to the FR12 DLA measurements
including velocity errors, in the same format as Fig.~\ref{fig:QSO_zerr}.
Here we assume no mean velocity shift, as there is nothing expected to
produce a systematic offset in DLA redshifts. Our global best fit yields
an RMS velocity error of $252^{+63}_{-53}\ \kms$ and an effective host halo
mass $M_{12} = 0.69^{+0.16}_{-0.14}$, with corresponding $b(\bar{M}) =
2.66\pm0.14$ at $z=2.5$. The inclusion of velocity errors makes an important
difference in the $\sigma = 1-4\ \mhmpc$ bin and little difference in other
bins. Our model of Gaussian velocity errors with a single RMS for DLAs is
questionable, as there is likely a strong dependence of the error on the
detection of associated metal lines, but it appears to yield a reasonable
fit to the data nonetheless. The $\chi^2$ of the global fit is 123 for 128
data points, with two fitting parameters. At $z=2.3$, our best-fit halo mass 
corresponds to a bias $b(\bar M)=2.39^{+0.13}_{-0.11}$. Our inferred bias and halo mass
are higher than those of the FR12 linear theory fit, which yields $\bdla =
2.17 \pm 0.2$ and corresponding $M_{12} = 0.45$, but the two determinations
are consistent at the $\sim 1\sigma$ level. The FR12 fit did not include
velocity errors, and it excluded points with $r<5\ \mhmpc$. Red curves in
Fig.~\ref{fig:DLA_zerr} show the prediction for $M_{12}=0.45$ and our
best-fit velocity error, and one can see that the differences from our
best-fit model prediction are small compared to the observational error
bars.

Analysis of the full BOSS DR12 data set will allow more detailed and
more precise measurements for the cross-correlations of both quasars
and DLAs, allowing narrower redshift bins for evolution measurements,
dividing samples into subsets of luminosity, equivalent width,
or metal-line strength, and achieving good statistical constraints
at large separations. Modeling these measurements with both linear
theory and non-linear predictions from LyMAS should yield numerous
insights into the quasar and DLA populations, and comparisons
of these model fits will either remove or sharpen the mild tensions
found here for the quasar cross-correlations. With the DR12 data, 
it may be possible to infer the RMS velocity of DLAs within their host 
halos by matching the small-scale correlation for the subset of DLAs that have 
metal-line absorption and thus small redshift measurement errors. This 
measurement would be an interesting new constraint on the physics of
DLA absorption.

\section{Conclusions}
\label{sec:con}

Cross-correlation with the densely mapped \lya\ forest opens a new window on 
the properties of quasars, DLAs, and potentially other populations of 
high-redshift objects. At large scales, one expects these correlations to be 
described by linear perturbation theory and linear bias, and precise measurements 
of quasar and DLA bias can be translated (for a specified cosmological model) 
to precise constraints on the masses of their host dark matter halos (FR12, FR13). 
Interpreting small scale cross-correlations requires a fully non-linear 
hydrodynamic model of the \lya\ forest. However, cosmological hydrodynamic 
simulations with the required resolution cannot model large enough volumes 
to provide accurate predictions of 3-d cross-correlations.

LyMAS provides a way to bootstrap hydrodynamic simulation results onto large 
volume $N$-body simulations and thus predict halo-flux cross-correlations over 
the full range of observable scales. When applied to the dark matter distribution 
of the Horizon-AGN or Horizon-noAGN simulations, LyMAS reproduces the 
cross-correlations \xips\ predicted using the full hydrodynamic \lya\ forest 
spectra almost exactly (Fig.~\ref{fig:LyMAS_vs_hydro}). The LyMAS predictions 
are insensitive to the dark matter mass resolution or smoothing scale over the 
range investigated here (Fig.~\ref{fig:resolution}). Furthermore, the predicted 
\xips\ is almost completely insensitive to AGN feedback as implemented in 
Horizon-AGN (Fig.~\ref{fig:AGN}), implying that this measure of cosmological 
structure is insensitive to feedback assumptions even at scales $\sigma=1-4$ \hmpc.

We have applied LyMAS to a $2048^3$-particle $N$-body simulation of a 1 \hgpc\ 
box, with the same WMAP7 cosmological parameters used for the Horizon simulations, 
to predict \xips\ for halo masses in the range \Mtwel\ $\approx0.5-50$ at $z=2.5$. 
Our numerical results (Fig.~\ref{fig:1000_fitted}) can be described by the 
empirically chosen Lorentzian form of equation (\ref{eq:Lorentz}), which can be 
used along with equations (\ref{eq:depth}) and (\ref{eq:FWHM}) and the 
parameters in Tables~\ref{tab:mass_params} and~\ref{tab:delta} to reproduce 
our results for any of the FR12/FR13 $\sigma$ bins and any halo mass in this 
range. The difference between the Lorentzian fit and the numerical results is 
significant at the level of our simulation's tiny statistical errors, but it is 
small compared to current observational errors. Except for the $\sigma=1-4$ 
\hmpc\ bin, the trend of cross-correlation amplitude with halo mass follows the 
standard large scale $b(M_h)$ relation from T10 (Fig.~\ref{fig:depvsmass}). 
Furthermore, even on small scales, \xips\ for a halo population with a range of 
masses depends only on the number-weighted halo bias (eq.~\ref{eq:bias_eff}). 
Modeling quasar or DLA observations down to these scales therefore improves 
constraints on the characteristic halo mass but does not provide information 
about the width or shape of the halo mass distribution.

We have compared the LyMAS results to linear theory predictions using bias 
parameters $b_F$ and $\beta_F$ for the \lya\ forest determined by fitting the 
LyMAS flux auto-correlation, and T10 values for the halo bias. We find 
surprisingly good agreement in \xips\ even at $\sigma=1-4$ \hmpc, with maximum 
deviations of $\sim10\%$ near $\pi=0$ \hmpc\ (Fig.~\ref{fig:cross_theory}). 
However, in a monopole-quadrupole representation, linear theory and LyMAS differ 
significantly in the quadrupole at $r<15$ \hmpc, while agreement for the 
monopole is good down to $r=3$ \hmpc\ (Fig.~\ref{fig:cross_monoquad}). In a 
$\xi(r,\mu)$ representation (Fig.~\ref{fig:cross_r5}), deviations are 
$\mu$-dependent, and they are larger for higher halo masses.

When comparing to the BOSS measurements of FR12 and FR13, we find that any 
choice of halo mass that fits \xips\ at $\sigma>7$ \hmpc\ overpredicts the 
amplitude of the cross-correlation at smaller scales (Fig.~\ref{fig:fitted_FR}). 
The most likely explanation for this discrepancy is scatter between halo 
velocities and measured redshifts in BOSS, which can arise from a combination 
of redshift measurement errors and random peculiar velocities of quasars or DLAs 
within their host halos; the velocities of the halos themselves are already 
accounted for in the LyMAS predictions. We fit the quasar (FR13) and DLA (FR12) 
measurements with a model that convolves the LyMAS predictions with a Gaussian 
distribution of redshift offsets of RMS amplitude $z_{\mathrm{err}}$, which 
removes the tension between large-scale and small-scale predictions. For quasars, 
we also incorporate the $-157$ km s$^{-1}$ mean redshift offset inferred by 
FR13 from the asymmetry of \xips.

For quasars (Fig.~\ref{fig:QSO_zerr}) we find $z_\mathrm{err}=399\pm21$ 
km s$^{-1}$, a magnitude that is compatible with other estimates of quasar 
redshift errors from BOSS spectra \citep{Paris2012}. Our best-fit halo mass is 
\Mtwel\ $=1.51^{+0.11}_{-0.10}$, which corresponds to a halo bias 
$b(\bar M)=3.18\pm0.06$ at $z=2.5$ for our WMAP7 cosmology (using the T10 halo bias 
relation). For DLAs (Fig.~\ref{fig:DLA_zerr}) we find 
$z_\mathrm{err}=252^{+63}_{-53}$ km s$^{-1}$, a value that appears high for pure 
measurement errors but may reflect a quadrature sum of measurement errors and 
halo internal velocities. Our best-fit halo mass is \Mtwel\ $=0.69^{+0.16}_{-0.14}$, 
with corresponding $b(\bar M)=2.66\pm0.14$ at $z=2.5$.

Relative to linear theory modeling, the LyMAS results have the advantage of 
fitting a fully non-linear model to the full range of measurements, and (as a 
necessary consequence) fitting for redshift scatter as well as effective halo 
mass. However, there are several caveats to our numbers. First, the slight 
mismatch between our simulation's $z=2.5$ redshift output and the $z=2.3$ 
central redshift of the BOSS measurements introduces ambiguity; our best-fit 
halo masses correspond to bias factors that are about 10\% lower at $z=2.3$ 
than at $z=2.5$ (see \S\ref{sec:obs}), and it is not clear which is the 
more appropriate characterization of our inferred halo bias. Second, the flux 
auto-correlation from the 1 \hgpc\ LyMAS box is significantly below that from 
the linear theory fit of \citet{Blomqvist2015} to the BOSS auto-correlation 
measurements at $z=2.3$ (Fig.~\ref{fig:auto}), and our best-fit values of 
$\beta$ and $b_F(1+\beta)$ do not agree with the values extrapolated from smaller 
volume hydro simulations by 
\citet{Arinyo2015}. This difference could reflect a combination of incorrect 
thermal structure of the diffuse IGM in our calibrating 
hydro simulations (e.g., if the real universe has a different history of helium 
reionization), inaccurate prediction of the large scale flux auto-correlation 
by LyMAS, uncertainties on scaling results between $z=2.3$ and $z=2.5$, and 
errors in the \citet{Blomqvist2015} measurements, which themselves require 
substantial corrections for the impact of continuum determination in the 
observed spectra.

Finally, for quasars, our fit has a statistically unacceptable $\chi^2$/d.o.f. 
$=290/160$. The largest contribution to $\chi^2$ comes from our innermost $
\sigma$-bin, where predictions are most sensitive to our assumption of Gaussian 
velocity errors. If we eliminate this bin from our mass fit (but retain the 
best-fit $z_\mathrm{err}$) we find \Mtwel\ $=2.19^{+0.16}_{-0.15}$, with 
corresponding $b(\bar M)=3.63^{+0.08}_{-0.07}$ at $z=2.5$. We regard this as our 
most reliable estimate of halo mass and halo bias for BOSS quasars because of 
its lower sensitivity to our $z_\mathrm{err}$ model. The $\chi^2$/d.o.f. improves 
to 197/142, but it is still high, reflecting a tension 
between lower masses favored by the $\sigma=4-15$ \hmpc\ data and the higher 
masses favored by $\sigma=15-60$ \hmpc\ data (Fig.~\ref{fig:QSO_chisq}). If 
instead we eliminate all data points with $r<15$ \hmpc\ but retain our best-fit 
$z_\mathrm{err}=399$ km s$^{-1}$, we obtain a best-fit halo mass of 
\Mtwel\ $=2.69^{+0.33}_{-0.35}$, corresponding to $b(\bar M)=3.85^{+0.14}_{-0.16}$ 
at $z=2.5$. The $\chi^2$/d.o.f. for this fit remains high at 221/132. For 
DLAs, the measurement errors are larger, and our fit (with \Mtwel\ 
$=0.69^{+0.16}_{-0.14}$ and $b(\bar M)=2.39^{+0.13}_{-0.11}$) has an acceptable 
$\chi^2$/d.o.f. $=123/126$.

Our DLA halo bias is consistent with that inferred by FR12 using linear theory 
modeling at separations $r>5$ \hmpc. Our quasar halo bias is lower than that 
inferred by FR13; an exact comparison is complicated by the slight difference 
in model redshift, but the best-fit FR13 parameters correspond to a predicted 
cross-correlation that is stronger by $\approx24\%$ (see Fig.~\ref{fig:QSO_zerr}). 
The main source of difference appears to be the selection of fitting range, 
as FR13's elimination of $r<15$ \hmpc\ separations removes the points that are 
favoring lower bias in our fit. The surprisingly good agreement between LyMAS 
and linear theory predictions of \xips\ suggests that smaller separations can 
be retained even in a linear theory fit, provided one accounts for quasar 
redshift errors.

Analyses of the final BOSS data set will allow cross-correlation measurements 
in bins of redshift and tracer properties (quasar luminosity, DLA metallicity) 
while retaining useful statistical precision on large scales. Comparison of 
observations, linear theory fits, and LyMAS modeling across these bins will 
allow tests for internal consistency (e.g., expected redshift scaling) and 
investigations of the relation between tracer properties and halo mass. The 
importance of $z_\mathrm{err}$ in our fits highlights a new opportunity to 
constrain the velocity distribution of quasars or DLAs within their host halos. 
For quasars, one can imagine measuring the mean offset and dispersion of 
redshifts inferred from different emission lines as a diagnostic of gas motions 
and ionization states in the immediate surroundings of the central black hole. 
Ionizing radiation from quasars should also affect cross-correlations on small 
scales (the ``transverse proximity effect"), and it may be possible to separate 
this impact from redshift dispersion effects via the expected dependence on 
quasar luminosity. For DLAs the comparison of RMS velocities to halo circular 
velocities may be a useful diagnostic for the kinematics of the metal-line 
absorption, perhaps distinguishing rotating disks from more chaotic gas 
distributions. The first studies of \lya\ forest cross-correlations have already 
yielded the most precise measurements of quasar and DLA clustering strength, 
and this tool holds considerable promise as a probe of the physics of 
high-redshift systems.

\section*{Acknowledgments}

We are grateful to numerous colleagues in the BOSS Collaboration for
fruitful discussions on this project, especially Jordi Miralda-Escud\'e.
We acknowledge support from the ``Programme National Cosmologie et Galaxies."
This work was supported in part by NSF Grant AST-1516997.

\appendix
\section{Cross-correlation Fitting Function}

\begin{figure*}
\begin{minipage}{175mm}
\centering
\includegraphics[width=\linewidth]{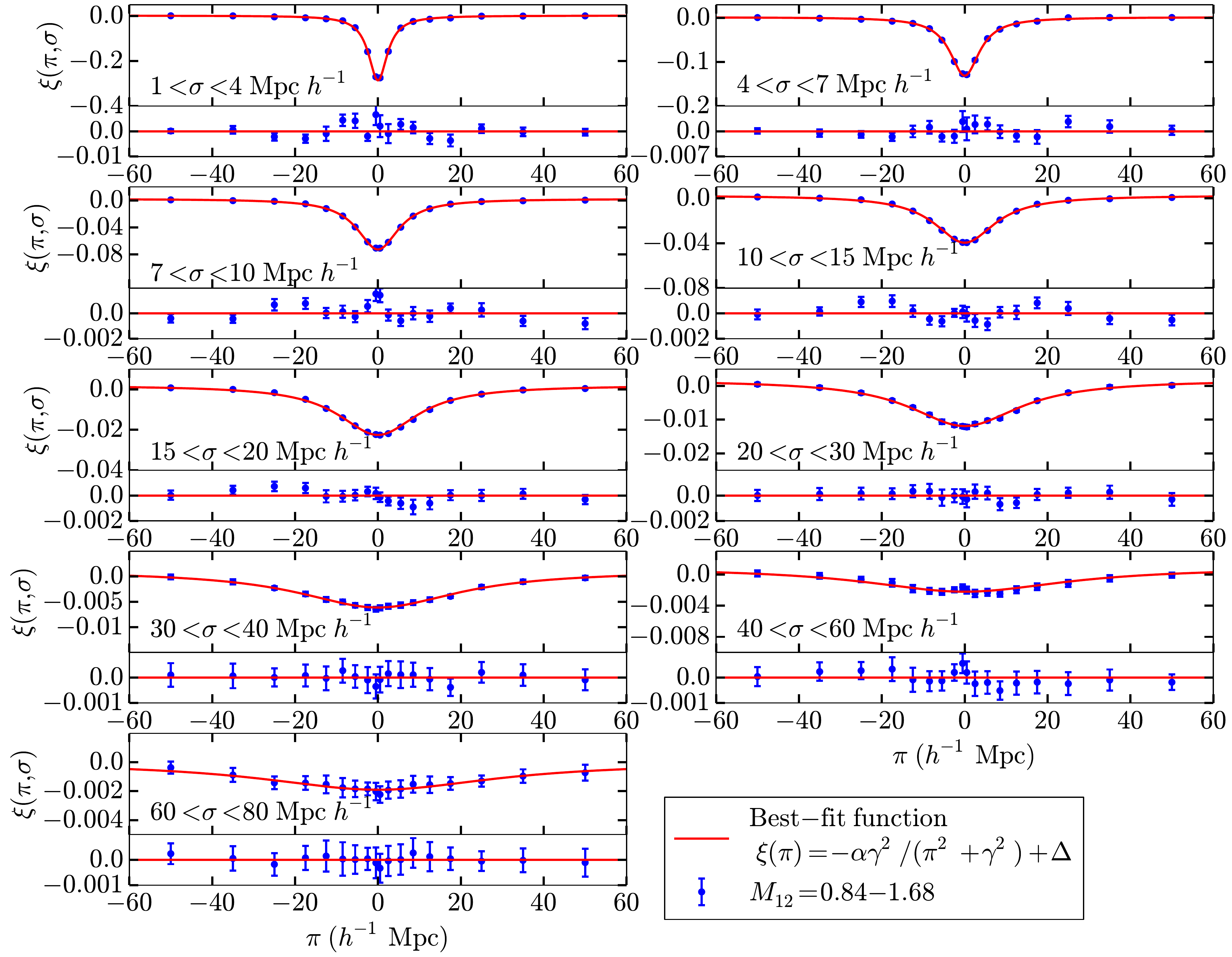}
\caption{The cross-correlation between dark matter halos and the \lya\
forest flux from applying LyMAS to the $2048^3$ (1 \hgpc)$^3$ dark matter
simulation with $r_s=0.5$ \hmpc\ for halos in the range \Mtwel\
$=0.84-1.68$ (blue points, with error bars estimated from 16 subvolumes).
Red curves show the best-fit Lorentzian fitting function $\xi(\pi)=-\alpha
\gamma^2/(\pi^2+\gamma^2)+\Delta$. Panels show different transverse
separation bins, increasing left to right and top to bottom, and residuals
are plotted beneath each panel.}
\label{fig:1000_fitted}
\end{minipage}
\end{figure*}

Fig.~\ref{fig:1000_fitted} shows the LyMAS predictions for the
cross-correlation of the \Mtwel\ $=0.84-1.68$ halos from the $2048^3$
simulation of the (1 \hgpc)$^3$ volume with $r_s=0.5$ \hmpc, for transverse
separation bins that range from $\sigma=1-4$ \hmpc\ up to $\sigma=60-80$
\hmpc. The correlation at $\pi=0$ weakens steadily with increasing
$\sigma$ as expected, from $|$\xipzero$|\approx 0.3$ in the $\sigma=1-4$
\hmpc\ bin to $0.04$ at $\sigma=10-15$ \hmpc\ to $0.002$ at $\sigma=60-80$
\hmpc. Results for other halo mass bins are qualitatively similar, with
the expected trend of stronger correlations for higher mass halos. Error
bars on these points are estimated by the subvolume method. Curves in
Fig.~\ref{fig:1000_fitted} show the least-squares fit of equation~\ref{eq:Lorentz} 
in each $\sigma$ bin,
where we have used the subvolume error bars (treated as diagonal) to fit
$\alpha$, $\gamma$, and $\Delta$. Formally the fits are often not
acceptable, but this may be a consequence of ignoring error covariances
and/or underestimated errors from our subvolume method.

\begin{figure*}
\begin{minipage}{175mm}
\centering
\includegraphics[width=\linewidth]{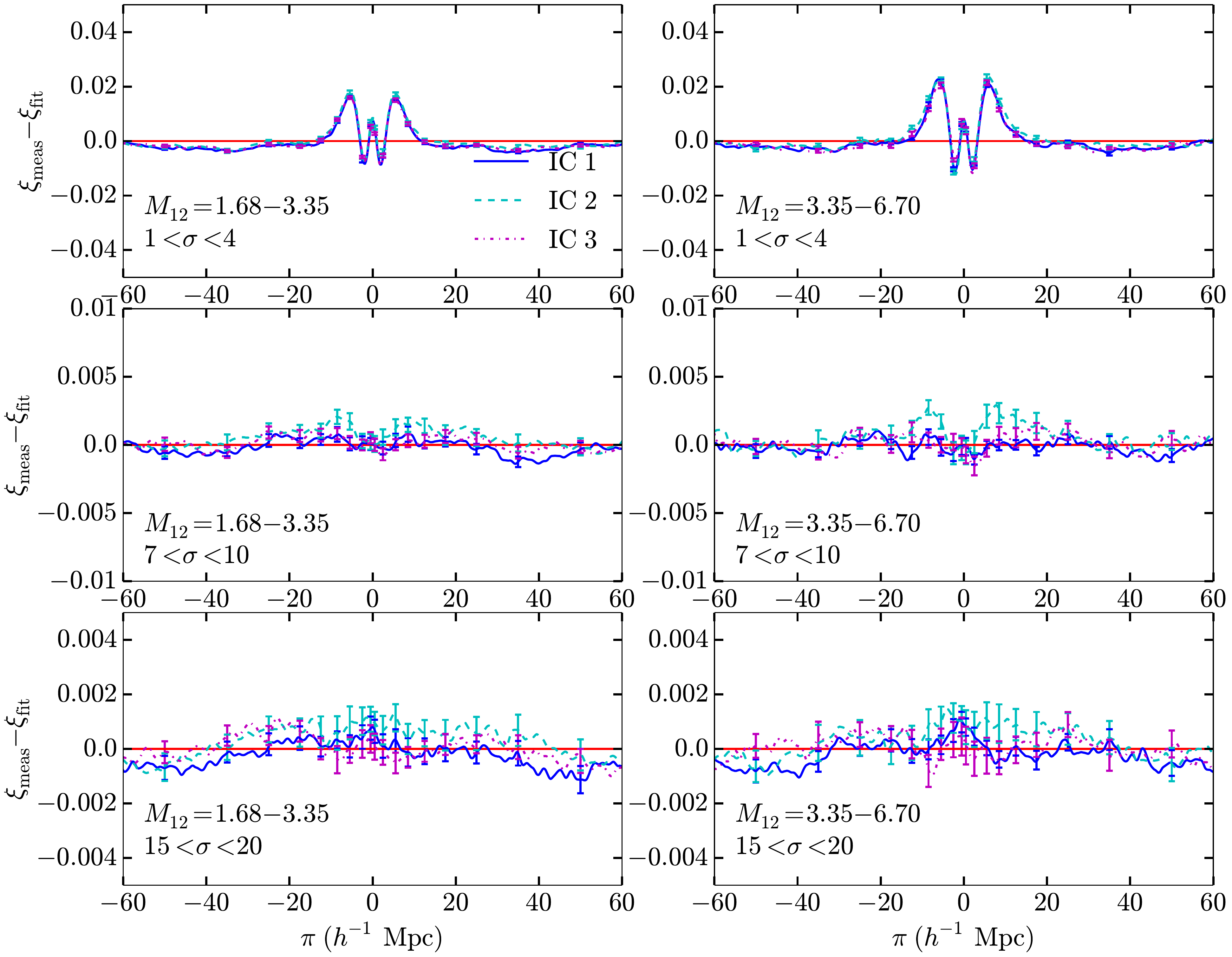}
\caption{The residuals of the halo-flux cross-correlation from fitting a Lorentz 
profile to the simulation marked ``IC 3." The three curves show residuals from 
three $1024^3$-particle (1 \hgpc)$^3$ dark matter simulations with $r_s=1.0$ \hmpc\ 
and different sets of initial conditions for the dark matter density field (blue solid, 
cyan dashed, and magneta dot-dashed lines). Rows show different
transverse separation bins $1<\sigma<4$ \hmpc, $7<\sigma<10$ \hmpc, and
$15<\sigma<20$ \hmpc, and columns show different halo mass bins \Mtwel\
$=1.68-3.35$ and \Mtwel\ $=3.35-6.70$. Error bars are computed by applying
the subvolume method to each simulation individually.}
\label{fig:cosvar}
\end{minipage}
\end{figure*}

To test if the subvolume method is an appropriate measure of cosmic
variance on the scale of the full volume, we compare three simulations with
different initial conditions for the dark matter density field, each (1
\hgpc)$^{3}$ in size, with $1024^3$ particles and a smoothing scale of
$r_s=1.0$ \hmpc. Fig.~\ref{fig:cosvar} shows the residual of the measured
cross-correlations ($\xi_{\mathrm{meas}}-\xi_{\mathrm{fit}}$) in each
simulation from a Lorentz profile fit to the simulation labeled ``IC 3,"
which has the same dark matter density field initial conditions as our
standard $2048^3$ simulation. We plot error bars on each simulation curve
computed by applying the subvolume method to that simulation. In the
smallest transverse separation bins, the difference among the three curves
is significantly larger than the subvolume error bars, but for
$\sigma=7-10$ \hmpc\ and larger the run-to-run variations are comparable in
size to the error bars. This means the subvolume method underestimates the
errors expected from cosmic variance in the smallest separation bins, but
for larger separations the subvolume errors are reasonable. Because we do
not have many 1 \hgpc\ simulations available, we use the subvolume errors
for our analysis.

Tables~\ref{tab:mass_params} and~\ref{tab:delta} can be used together
with equations~(\ref{eq:Lorentz}), (\ref{eq:depth}), and (\ref{eq:FWHM})
to accurately reproduce the $z=2.5$ LyMAS predictions of $\xi(\pi,\sigma)$
for any value of $M_h$, with typical residuals comparable to those 
shown in Fig.~\ref{fig:1000_fitted}. They thus provide a compact quantitative
summary of our numerical results. Table~\ref{tab:delta} gives the mean value of $\Delta$ in each
halo mass bin, averaged over the transverse separation bins. There does
not appear to be any significant trend in $\Delta$ with transverse
separation, and only a weak trend with halo mass. Nevertheless, we include
it in our model, using the average values recorded in the table.

\begin{table}
\centering
\caption{Best-fit parameters for fits to the trend
in halo-flux cross-correlation strength with halo mass $M_h$ and transverse
separation $\sigma$ bin, $|\xi(\pi=0,\sigma)|=A(M_h/4\times10^{12}\ h^{-1}\
M_\odot)^m$. Errors on parameters are from the subvolume method.}
\begin{tabular}{c c c} \hline
$\sigma$ bin (\hmpc) & log $A$ & $m$ \\
\hline
$1-4$ & $-0.45\pm0.003$ & $0.215\pm0.005$ \\
$4-7$ & $-0.75\pm0.003$ & $0.280\pm0.007$ \\
$7-10$ & $-1.00\pm0.004$ & $0.255\pm0.013$ \\
$10-15$ & $-1.25\pm0.006$ & $0.270\pm0.012$ \\
$15-20$ & $-1.50\pm0.008$ & $0.260\pm0.015$ \\
$20-30$ & $-1.80\pm0.011$ & $0.275\pm0.025$ \\
$30-40$ & $-2.10\pm0.020$ & $0.280\pm0.025$ \\
$40-60$ & $-2.45\pm0.026$ & $0.300\pm0.038$ \\
$60-80$ & $-2.80\pm0.036$ & $0.240\pm0.035$ \\
\hline
\end{tabular}
\label{tab:mass_params}
\end{table}

\begin{table}
\centering
\caption{The $\Delta$ value reported here is the
mean across $\sigma$ bins in the corresponding halo mass bin, since we do
not detect any significant trend in $\Delta$ with $\sigma$ bin. Errors on
parameters are from the subvolume method.}
\begin{tabular}{c c} \hline
\Mtwel\ bin & $\Delta$ \\
\hline
$0.42-0.84$ & $0.0012\pm0.0003$ \\
$0.84-1.68$ & $0.0011\pm0.0003$ \\
$1.68-3.35$ & $0.0014\pm0.0003$ \\
$3.35-6.70$ & $0.0014\pm0.0003$ \\
$6.70-13.40$ & $0.0015\pm0.0003$ \\
$13.40-26.80$ & $0.0019\pm0.004$ \\
$26.80-53.60$ & $0.0019\pm0.0004$ \\
\hline
\end{tabular}
\label{tab:delta}
\end{table}

\end{document}